\documentclass[11pt, a4paper,twoside]{article}
\usepackage{amssymb,amsmath,amsthm,latexsym,bbm,mathrsfs,nicefrac,extpfeil,dsfont,booktabs,caption, multicol,float,extarrows}
\usepackage{a4wide,microtype,xcolor} 
\usepackage{hyperref} 
\usepackage[latin1]{inputenc} 
\usepackage{natbib} 
\usepackage{graphicx} 
\bibpunct{(}{)}{;}{a}{,}{,} 
\graphicspath{{./Grafiken/}} 

\newtheorem{theorem}{Theorem}[section]

\newtheorem{definition}[theorem]{Definition}
\newtheorem{lemma}[theorem]{Lemma}
\newtheorem{corollary}[theorem]{Corollary}

\newcommand{\E}{\mathbb{E}}
\newcommand{\V}{\mathbb{V}\!{ar}}

\newcommand{\niton}{\not\owns}

\title{Multi-scale detection of variance changes in renewal processes in the presence of rate change points}
\author{{\sc Stefan Albert$^{1,*}$} \and {\sc Michael Messer$^{1,*}$} \and {\sc Julia Schiemann$^2$} \and {\sc Jochen Roeper$^3$} \and {\sc Gaby Schneider$^1$}\\[1.5ex]
	{$^1$Institute of Mathematics, Goethe University, Frankfurt (Main), Germany}\\
	{$^2$ Centre for Integrative Physiology, University of Edinburgh, UK}\\
	{$^3$ Institute of Neurophysiology, Goethe University, Frankfurt (Main), Germany}\\[1.5ex]{$^*$ equal contribution}\\[1.5ex]{Corresponding author: Gaby Schneider, schneider@math.uni-frankfurt.de}
}
\date{}

\begin{document}
\maketitle
\newpage

\begin{abstract}
Non-stationarity of the rate or variance of events is a well-known problem in the description and analysis of time series of events, such as neuronal spike trains. A multiple filter test (MFT) for rate homogeneity has been proposed earlier that detects change points on multiple time scales simultaneously. It is based on a filtered derivative approach, and the rejection threshold derives from a Gaussian limit process $L$ which is independent of the point process parameters. 

Here we extend the MFT to variance homogeneity of life times. When the rate is constant, the MFT extends directly to the null hypothesis of constant variance. In the presence of rate change points, we propose to incorporate estimates of these in the test for variance homogeneity, using an adaptation of the test statistic. The resulting limit process shows slight deviations from $L$ that depend on unknown process parameters. However, these deviations are small and do not considerably change the  properties of the statistical test. This allows practical application, e.g.~to neuronal spike trains, which indicates various profiles of rate and variance change points.
\end{abstract}


\section{Introduction}
\paragraph{Motivation}
Non-stationarity of the rate or variance of events is a well-known problem in the description and analysis of time series of events. For example, neuronal activity patterns in spike trains are often analyzed with point process models \citep{dayan,kass2005,gruen_para}. As such analyses can be affected by changes in process parameters, it is often necessary to use preprocessing steps that divide the processes into sections with approximately constant parameters \citep{gruen2002,schneider2008,staude2010,quiroga2013}. These preprocessing steps use models with step functions for the parameters and aim at detecting the points in time when the parameters change, i.e., the change points.

For the detection of change points in the rate, several techniques have been developed, e.g., \cite{bertrand2000,lavielle2000, bertrand2011,horvath2008,kirch2014}. Interesting multi scale methods have been proposed by \cite{fryz2014,matteson2014,messer_paper,munk2014}. For a general survey about change point methods we refer to the books \cite{basseville,brodsky,csorgo} or the review article of \cite{aue2013}. 

Some of these techniques can also be applied to the detection of variance change points, and other approaches to the analysis of variance homogeneity have been proposed by \citet{hsu1977, inclan1993, inclan1994,chen1997, whitcher2000, killick2010, zhao2010, noorossana2012, Killick2013, Nam2015, korkas2016}.
However, most available methods use specific assumptions on the underlying distribution, e.g., Gaussian sequences, or aim at detecting at most one change point. In addition, they usually assume the rate to be constant. Few applied approaches simultaneously deal with potential rate and variance changes \citep{hawkins2005,rodionov2005}. Recently, \citet{dette2015} proposed a statistical test for the null hypothesis of constant variance in the presence of a smoothly varying mean. 

\begin{figure}[htb]
          	\centering
          	\includegraphics[clip=TRUE,scale=0.5,trim= 0 0em 0 0em]{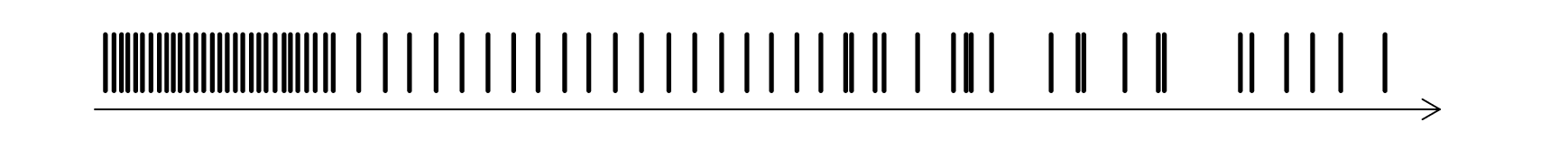}
		\caption{A point process with non-stationary rate and variance profile.}
						\label{fig_example_variancep}
\end{figure}

Here we aim at detecting both, rate and variance change points (see~Figure \ref{fig_example_variancep}) that may occur in multiple time scales. To that end, we propose a two-step procedure that first tests the null hypothesis of rate homogeneity in the potential presence of variance changes and estimates change points in the rate if the null hypothesis is rejected. In the second step, we test the null hypothesis of variance homogeneity and estimate variance change points. As this step requires estimation of the underlying rate, we propose to plug in the estimated rate change points derived in the first step. Our procedure  is applicable to renewal processes with a wide range of life time distributions. In our setting we are not restricted to the alternative of at most one change-point and allow for change points in the rate such that the rate is given by a step function.

\paragraph{The multiple filter test (MFT)}
The procedure builds on a recently proposed multiple filter test (MFT) \citep{messer_paper} and a corresponding algorithm (MFA). These were originally designed for the detection of change points in the rate on different time scales in a wide class of point processes which allows for a certain variability in the variance of the life times and are thus considered suitable candidates for the first step of rate change detection here. They will be modified in the second step to allow for the detection of variance changes. The filtered derivative idea used in the MFT works as follows. Assume that each life time $\xi_i$ depends on a parameter $\vartheta_i$ of which change points are to be detected. 
For each time $t$, compare the information about $\vartheta$ in the left and right window of size $h$ denoted here by $J_{\text{le}}:=J(t-h,t]$ and $J_{\text{ri}}:=J(t,t+h]$, using a scaled process $G:=(G_{h,t})_{t}$ with
\begin{equation}
\label{def_ght}
G_{h,t}:=\frac{J_{\text{ri}}-J_{\text{le}}}{\hat s_t},
\end{equation}
where $\hat s$ is an appropriate estimator for the standard deviation of the numerator under the null hypothesis of no change point in $\vartheta$. For example, \citet{messer_paper} use the numbers of events in the left and right window in order to detect change points in the mean of the life times. Under mild assumptions, the process $G$ converges weakly under the null hypothesis to a process  $L:=(L_{h,t})_t$ given by
\begin{equation}
\label{def_lht}
L_{h,t}:=\frac{(W_{t+h}-W_t)-(W_t-W_{t-h})}{\sqrt{2h}},
\end{equation}
for a standard Brownian motion $(W_t)_{t\ge0}$, if the window size $h$ grows linearly with the total time $T$. Note that the process $L$ does not depend on the parameters of the point process.

While under the null hypothesis $G$ fluctuates around zero, a change  in $\vartheta$ at time $t$ should cause systematic deviations from zero. Therefore, a large temporal maximum $M_h:=\max_t |G_{h,t}|$ indicates a change point in $\vartheta$. Using a finite set of multiple windows $H=\{h_1,   \ldots,  h_k\} \subset (0,T/2]$ simultaneously, the MFT allows for the detection of change points on different time scales. The global maximum $M$ of all processes $\left(G_{h,t}\right)_{h\in H}$  serves as a test statistic whose distribution can be approximated from the corresponding limit processes $\left(L_{h,t}\right)_{h\in H}$, i.e.,
\begin{equation}
\label{def_m}
M:=\max_{h \in H} M_h = \max_{h \in H} \max_{t \in [h,T-h]} | G_{h,t} | \sim \max_{h \in H} \max_{t \in [h,T-h]} | L_{h,t} |.
\end{equation}
By simulating these limit processes $\left(L_{h,t}\right)_{h\in H}$ as functionals of the same underlying Brownian motion, the rejection threshold $Q$ of the MFT can be obtained. We stress that the derivation of the quantile $Q$ works in two steps: First, we use that the maximum of all processes $(G_{h,t})_t$ over all windows $h$ converges to the maximum of the limit processes $(L_{h,t})_t$ over all windows $h$, the latter being a functional of a standard Brownian motion and particularly independent from parameters of the input spike train.  Then, in a second step, we simulate $Q$ as a quantile of the limit law. (To the best of our knowledge, there is no closed formula expression for the  limit law where we could directly read $Q$ from.) The main reason for this two step approach is that it allows for the simultaneous application of multiple windows, which helps to improve the detection of change points that appear on different time scales: small windows are more sensitive to frequent change points, while larger windows have higher power and thus improve the detection of parameter changes of smaller magnitude. The MFT yields comparable results to other change point methods (for an example see Table \ref{tab:CompMFTPLS} in Appendix \ref{sect:comparison}).

\paragraph{The MFT for variance changes -- outline}
In order to perform the second step of  change point detection in the variance, we extend the MFT here, where now the relevant information $J$ in the process $G$ from (\ref{def_ght}) is an estimator of $\sigma^2$ (section \ref{subsec_meanconstant}),
 \begin{equation}
\label{eq_def_ght_int}
G_{h,t}:=\frac{\hat\sigma_{\text{ri}}^2-\hat\sigma_{\text{le}}^2}{\hat s_t}, 
\end{equation}
where $\hat s_t$ denotes an estimator of the standard deviation of the numerator. Assuming first rate homogeneity with i.i.d.~life times, we show that under the null hypothesis of constant variance, $G$ converges weakly in Skorokhod topology to the same limit process $L$ (eq.~(\ref{def_lht})) (section \ref{sect:Th23}). This enables to test for and estimate change points in the variance analogously to rate change points, applying the modified process $G$ from equation (\ref{eq_def_ght_int}). We then deal with processes that contain rate and variance changes by investigating  the impact of one rate change point on the limit behavior of $G$ in section \ref{subsec_limit}. Under the null of constant variance, the limit process is a continuous, $2h$-dependent zero-mean, unit-variance Gaussian process $\widetilde L$ similar to $L$ (Theorem \ref{main_theorem_b}), with slight changes in the covariance structure in the neighborhood of a rate change point. 
As the process $\widetilde L$ depends on unknown point process parameters, we suggest to use $L$ to derive the rejection threshold of the test. This is supported by our theoretical and simulation results.

The practical performance of the MFT and the corresponding MFA \citep{messer_paper} for the detection of variance change points is presented in section \ref{sec_eval}. Our simulations suggest that the significance level of the MFT for variance changes is kept for a wide range of parameter settings also in cases with additional rate changes. Further, the detection probability of variance change points is barely affected by the necessity to estimate rate change points and depends on the magnitude of the variance change as well as on the regularity of the process. We  present an example for the MFA on rate and variance change point detection and illustrate the importance of including existing rate change points in the estimation of variance change points. Finally, we use the MFA in section \ref{sec_data} to estimate rate and variance change points in spike train recordings obtained in the substantia nigra of anaesthetized mice.

\section{The MFT for testing variance homogeneity}
\label{sec_notandmft}

Here we derive the limit distribution of the filtered derivative process $G$ when testing for variance homogeneity. The rejection threshold of the statistical test can be obtained as described in the introduction by simulation of the respective functional of the limit process. We first define the notation and model assumptions. Section \ref{subsec_meanconstant} then elaborates on the explicit structure of $G$ when testing for variance homogeneity. Limit results for $G$ under constant rate and under one change point in the rate are given in sections \ref{sect:Th23} and \ref{subsec_limit}, respectively.\\ 

Throughout this paper we let $\Phi$  denote a point process on the positive line with events $S_1,S_2,\ldots$ and life times $(\xi_i)_{i \geq 1}$ with $\xi_1=S_1$ and $\xi_i=S_i-S_{i-1}$, $i=2, 3, \ldots$ First we define a class $\mathscr{R}$ of renewal processes on the positive line with $\xi_i \in \mathscr{L}^{4}$ (Definition \ref{def_renewal}). The models with change points in the mean and/or variance considered here are then given as piecewise elements of $\mathscr{R}$ (Definition \ref{def_fullmodel}).

\begin{definition}{(Renewal process)} 
	\label{def_renewal}	
The class of point processes with i.i.d., a.s.~positive life times $(\xi_i)_{i \geq 1}$ with $\xi_1\in\mathscr{L}^{4}$ is called $\mathscr{R}$. 	
\end{definition}

\begin{figure}[htb]
          	\centering
          	\includegraphics[clip=TRUE,scale=0.5,trim= 0 0em 0 0em]{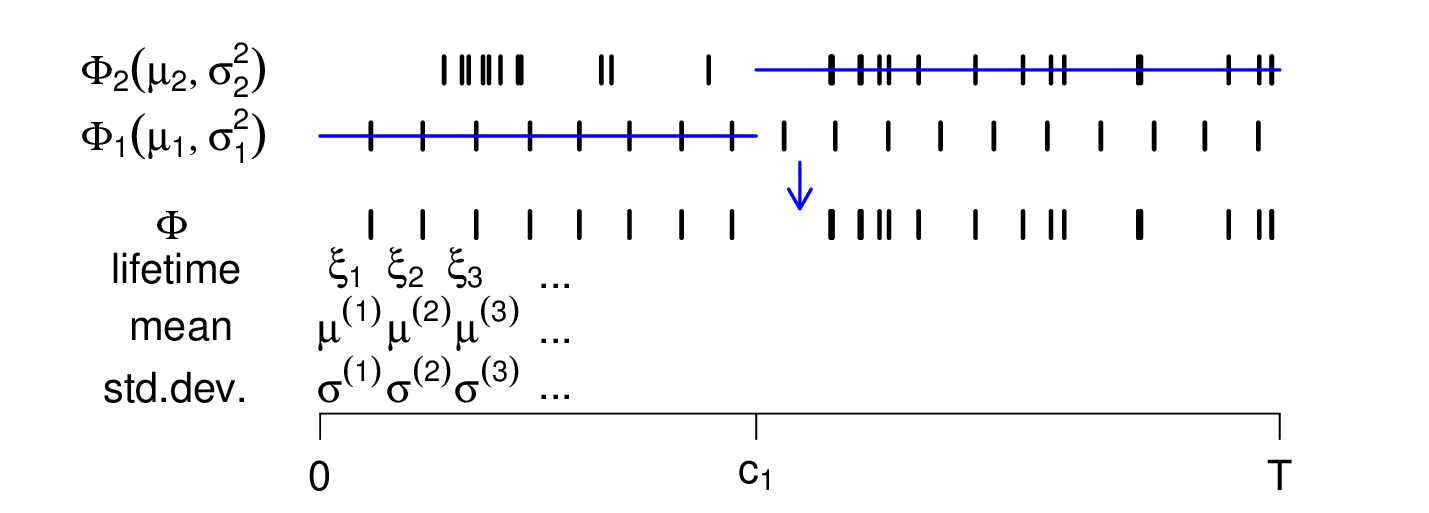}
		\caption{A realization of a process $\Phi$ according to Definition \ref{def_fullmodel}. $\Phi$ originates from two processes $\Phi_1(\mu_1,\sigma_1^2)$ and $\Phi_2(\mu_2,\sigma_2^2)\in\mathscr R$. Each life time $\xi_i$ has mean $\mu^{(i)}$ and standard deviation $\sigma^{(i)}$.}
						\label{fig_defg}
\end{figure}

A process $\Phi\in \mathscr{R}$ whose life times have mean $\mu$ and variance $\sigma^2$ and $\nu^2:=\mathbb Var((\xi_1-\mu)^2)$ is therefore denoted by $\Phi(\mu,\sigma^2) := \Phi(\mu,\sigma^2,\nu^2)$. The inverse of the mean, $\mu^{-1}$, is termed the rate of $\Phi$. A class of processes that are piecewise  elements of $\mathscr{R}$ is used in order to introduce rate and/or variance changes.
\begin{definition} (Renewal process with change points in the mean or variance)\label{def_fullmodel}\\ 
For $T>0$ let $C:= \{c_1, \ldots, c_k \}$ be a set of change points with $0<c_1 < \ldots < c_k <T$. At time $t=0$ start $k+1$ independent elements of $\mathscr{R}$
	\begin{equation} 
		\label{construct_processes}
		\Phi_1\left(\mu_1,{\sigma_1^2}\right), \ldots, \Phi_{k+1}\left(\mu_{k+1},\sigma_{k+1}^2\right),
	\end{equation}
	with $(\mu_{i},\sigma_{i}^2) \neq (\mu_{i+1},\sigma_{i+1}^2)$.	Let $c_0:=0$, $c_{k+1}:=T$ and define
	\begin{equation*}
	\Phi:= \bigcup_{j=1}^{k+1} \Phi_j | _{(c_{j-1},c_j]},
	\end{equation*}
	where $\Phi_j | _{(c_{j-1},c_j]}$ denotes the restriction of $\Phi_j$ to the interval $(c_{j-1},c_j]$. 
\end{definition}
The family of processes which derive according to Definition \ref{def_fullmodel} is called $\mathscr{M}$ (see Figure \ref{fig_defg} for an example).  
For $\Phi\in\mathscr M$, at each change point $c_i$ the rate and/or the variance changes, such that the rate and variance constitute step functions. 

Thus, we test the null hypothesis $H_0: \sigma_1^2=\ldots=\sigma_{k+1}^2$ against the alternative $H_A: \exists i,j: \sigma_i^2 \neq \sigma_j^2$, where we allow for an unknown number of potential additional change points in the rate that may or may not occur simultaneously with rate changes. Note that we require the mean in order to estimate the variances ($\hat \sigma^2_{\rm{ri}}, \hat \sigma^2_{\rm{le}}$) and to derive the test statistic $G$. We therefore first formulate the theory without explicit assumptions on the mean, letting  $\mu^{(i)}$ denote the mean of each individual life time $\xi_i$. Later on we distinguish between the case with constant mean and the case where the mean follows a step function, and we investigate the behavior under estimation of $\mu^{(i)}$.

\subsection{Filtered derivative approach for the variances}
\label{subsec_meanconstant}
As explained in the introduction, we test the null hypothesis using a window size $h>0$ and the filtered derivative process from (\ref{eq_def_ght_int}) for $t \in \tau_h:=[h,T-h]$ defined as  
 \begin{equation*}
 G_{h,t}:=\frac{\hat\sigma_{\text{ri}}^2-\hat\sigma_{\text{le}}^2}{\hat s_t}
 \end{equation*}
 if $\hat s_t >0$ and $G_{h,t}:=0$ otherwise. The numerator is given by the standard variance estimators (eq.~(\ref{eq_sigma2})), and $\hat s_t$ is a local estimator of the standard deviation of the numerator (eq.~(\ref{def_s})). We use the notation
\begin{equation}
\label{eq_def_vi}
V_i:=(\xi_i-\mu^{(i)})^2
\end{equation}
with $\mu^{(i)}:=\E[\xi_i]$, $(\sigma^{(i)})^2:=\E[V_i]=\V(\xi_i)$ and $(\nu^{(i)})^2:=\V(V_i)$ (Figure \ref{fig_defg}). In order to include estimated rates, we use an estimator $\hat\mu^{(i)}$ of $\mu^{(i)}$ and define the estimator of $V_i$ as
\begin{equation}
\label{eq_def_vi_dach}
\hat V_i:=(\xi_i-\hat\mu^{(i)})^2.
\end{equation}
As estimator $\hat \mu^{(i)}$ we later use a global estimator derived as the mean of all life times (Theorem~\ref{main_theorem_a}) or a local estimator derived between estimated change points (Theorem~\ref{main_theorem_b}).

If $\hat I_{\text{le}}$ and $\hat I_{\text{ri}}$ denote the sets of life times in $(t-h,t]$ and $(t,t+h]$ which do not overlap a rate change point, the standard variance estimators are given by
\begin{align}
\label{eq_sigma2}
\hat{\sigma}_{\text{le}}^2:=\frac{1}{|\hat I_{\text{le}}|}\sum\limits_{i \in  \hat I_{\text{le}}}^{}{\hat V_i} 
\quad \quad \text{and} \quad \quad 
\hat{\sigma}_{\text{ri}}^2:=\frac{1}{|\hat I_{\text{ri}}|}\sum\limits_{i \in  \hat I_{\text{ri}}}^{}{\hat V_i} 
\end{align}
if $|\hat{I}_{\text{ri}}|,|\hat I_{\text{le}}|>0$ and zero otherwise. The estimator $\hat s_t^2$ of $\V(\hat{\sigma}_{\text{ri}}^2-\hat{\sigma}_{\text{le}}^2)$ in the denominator of $G$ is defined as
\begin{equation}
\label{def_s}
	\hat{s}^2_{t}:=\frac{\hat{\nu}^2_{\text{ri}}}{h /\hat{\mu}_{\text{ri}}}+\frac{\hat{\nu}^2_{\text{le}}}{h /\hat{\mu}_{\text{le}}},
\end{equation}
where $\hat{\mu}_{\text{ri}}$ and $\hat{\mu}_{\text{le}}$ are the  means of the life times in $\hat I_{\text{ri}}$ and $\hat I_{\text{le}}$ and the numerators are estimated as
\begin{align}
\hat{\nu}^2_{\text{le}}:= \frac{1}{|\hat I_{\text{le}}|}\sum\limits_{i \in  \hat I_{\text{le}}}^{}{(\hat V_i-\hat \sigma_{\text{le}}^2)^2}
\quad \quad \text{and} \quad \quad 
\hat{\nu}^2_{\text{ri}}:= \frac{1}{|\hat I_{\text{ri}}|}\sum\limits_{i \in  \hat I_{\text{ri}}}^{}{(\hat V_i-\hat \sigma_{\text{ri}}^2)^2}
\label{def_nu_exact}
\end{align}
for $|\hat I_{\text{le}}|>0$ and $|\hat I_{\text{ri}}|>0$ and zero otherwise. This is motivated by the CLT $\sqrt{t/\mu}(\hat\sigma_t^2-\sigma^2)\stackrel{d}{\longrightarrow}N(0,\nu^2)$ as $t\to\infty$, where $\hat\sigma_t^2$ denotes the empirical variance of all life times up to time $t$.

\subsection{Limit behaviour of \texorpdfstring{$G_{h,t}$}{Ght} under a constant rate}\label{sect:Th23}
If the mean of the life times  is constant  $\mu$, one can show the following Theorem \ref{main_theorem_a}, which allows application of the multiple filter approach. We use the extended filtered derivative process $G^{(n)}:=(G^{(n)}_{h,t})_t:=(G_{nh,nt})_t$ from (\ref{eq_def_ght_int}) where the window size and the time grow linearly in $n$. Furthermore, we use the globally estimated mean $\hat{\mu}:=\hat{\mu}_{nT}:=(1/N_{nT})\sum_{i=1}^{N_{nT}}{\xi_i}$ as estimator for each $\mu^{(i)}$, where $N_t$ denotes the number of events up to time $t$.

\begin{theorem}
\label{main_theorem_a}
	Let $T>0$ and $h \in (0,T/2]$ be a window size.
	 If $\Phi\in\mathscr M$ with constant $\mu$ and $\sigma^2$ using the globally estimated mean $\hat\mu$ we have  in $(D[h,T-h],d_{SK})$  for $n \to \infty$
	\begin{equation*}
		G^{(n)} \xlongrightarrow[]d{} L, 
	\end{equation*}
	with $L$ as defined in (\ref{def_lht}).
	\end{theorem}

For a proof see Appendix \ref{sec_proof_a}. This assertion holds particularly for a constant and known mean, i.e., if $\hat\mu^{(i)}=\mu_1 \ \forall i \geq 1$. Note that by extension of Theorem \ref{main_theorem_a} one can analogously test the null hypothesis of homogeneity of the $k$-th order moments $m_k:=\E[\xi_1^k]$ for every fixed $k$ assuming constant lower order moments.

\subsection{Limit behavior of \texorpdfstring{$G$}{G} with one rate change point}

\label{subsec_limit}
In this section we extend Theorem \ref{main_theorem_a} allowing for one rate change point, while testing the null hypothesis of variance homogeneity. Assuming a process with at most one rate change point, the process $G$ can be shown to converge against a limit process $\widetilde L$ (Theorem \ref{main_theorem_b}), which is, like $L$, a zero-mean $2h$-dependent Gaussian process with unit variance. It differs from $L$ only in the covariance in the $3h$-neighborhood of a change point $c$ (see section \ref{subsec_signi_rinh} and Fig.~\ref{fig_l_comp} C,D).

\begin{theorem}
\label{main_theorem_b}
	Let $\Phi^{(n)}\in\mathscr M$ (Def.~\ref{def_fullmodel}) with at most one rate change and no variance change, as follows. Let $\Phi_1(\mu_1,\sigma_1^2,{\nu_1}^2), \Phi_2(\mu_2,\sigma_2^2,{\nu_2}^2) \in \mathscr R$ with $\mu_1 \neq \mu_2$, $\sigma_1^2 = \sigma_2^2$. For $c\in(0,T]$ and $n=1,2,\ldots$ let
	\begin{equation}
		\Phi^{(n)}:=\Phi_1| _{[0,nc]}+\Phi_{2}| _{(nc,nT]}, 
		\label{model_exp}
	\end{equation}
	meaning that $\Phi^{(n)}$ fulfills $H_0$. Assume a consistent estimator $\hat c$ of $c$ with
		  \begin{equation}
			\label{eq:assump}
		  |\hat c-c|=o_{\mathbb{P}}(1/n)
		  \end{equation} 
			where $o_\mathbb{P}(\cdot)$ is the small $o$-notation with respect to convergence in probability. Let $G^{(n)}$ be the filtered derivative process associated with $\Phi^{(n)}$ using the empirical means $\hat \mu_1^{\hat c}$, $\hat \mu_2^{\hat c}$ estimated in the intervals $[0,\hat c)$ and $[\hat c,T]$. Then with $\widetilde L$ from (\ref{def_l}), as $n \to \infty$, we have
	\begin{equation*}
		G^{(n)} \xrightarrow [\quad]{d} \widetilde L,
	\end{equation*}
	where $\stackrel{d}{\longrightarrow}$ denotes weak convergence in the Skorokhod topology. The marginals $\widetilde L_{h,t}$ of the limit process $\widetilde L$ equal $L$ outside the $h$-neighborhood of $c$ and are given by 
	{{
	\begin{equation}
	\label{def_l}
\widetilde L_{h,t}=		\begin{cases}
		L_{h,t}, &  |t-c|>h, \\
		\frac{\sqrt{{(\mu_{\rm{ri}}\nu_2)}^2 /(\mu_2 h^2)}(W_{t+h}-W_c)+\sqrt{{(\mu_{\rm{ri}} \nu_1)}^2/( \mu_1 h^2)}(W_{c}-W_t)-\sqrt{\mu_1{ \nu_1}^2/h^2}(W_t-W_{t-h})}{s^{(1)}_t}, &  t \in [c-h, c],\\
		\frac{\sqrt{\mu_2{ \nu_2}^2/  h^2}(W_{t+h}-W_t)-\sqrt{{(\mu_{\rm{le}} \nu_2)}^2/( \mu_2  h^2)}(W_t-W_c)-\sqrt{{(\mu_{\rm{le}}\nu_1)}^2 /(\mu_1 h^2)}(W_c-W_{t-h})}{s^{(1)}_t}, &  t \in (c, c+h],\\
		\end{cases}
	\end{equation}}}
for  a standard Brownian motion  $(W_t)_{t\ge 0}$. The functions $\mu_{\rm{ri}}:=\mu_{{\rm ri},h,t}, \mu_{\rm{le}}:=\mu_{{\rm le},h,t}$ are the limits of the empirical means $\hat\mu_{\rm{ri}}, \hat\mu_{\rm{le}}$ and are given by $\mu_{{\rm ri},h,t}:=\mu_1$ for $t \leq c-h$, $\mu_{{\rm ri},h,t}:=\mu_2$ for $t > c$ and
\begin{equation}
\label{def_muaverage2}
\mu_{{\rm ri},h,t}:=\frac{h}{(c-t) /\mu_1 +(t+h-c) / \mu_2},
\end{equation}
for $t\in (c-h,c]$ and analogously for $\mu_{\rm{le}}$. The true order of scaling $\left((s_t^{(n)})^2\right)_{t \in \tau_h}$ is defined by $\frac{2{\nu_1}^2}{nh/\mu_1}$ for $t< c-h$, by $\frac{2{\nu_2}^2}{nh/\mu_2}$ for $t> c+h$ and for $|t-c| \leq h$ by the following linear interpolation
	\begin{equation}
	\label{def_stheo}
	(s_t^{(n)})^2:=(s_{h,t}^{(n)})^2:=
	\begin{cases}
	\frac{1}{n}\left( \frac{\mu_1 {\nu_1}^2 }{h} + \frac{(c-t)}{h^2 \mu_1}{(\mu_{\rm{ri}}\nu_1)}^2+\frac{(t+h-c)}{h^2\mu_2}{(\mu_{\rm{ri}} \nu_2)}^2 \right)	,&  \text{if } c-h \leq t \leq c\\
	\frac{1}{n}\left(\frac{(c-(t-h))}{h^2 \mu_1}{(\mu_{\text{le}}\nu_1)}^2+\frac{(t-c)}{h^2\mu_2}{(\mu_{\rm{le}} \nu_2)}^2 + \frac{\mu_2 {\nu_2}^2 }{h}\right), &  \text{if } c < t \leq c+h.\\
	\end{cases}
	\end{equation}
\end{theorem}
The proof of Theorem \ref{main_theorem_b} can be found in  Appendix \ref{sec_proof_b}. 
Note that analogous results hold if there are several rate change points with pairwise distances each larger than $2h$. Furthermore, note that the proof of Theorem \ref{main_theorem_b} is based on a FCLT and a consistent estimator of $s_t$. Therefore the result can be shown not only for renewal processes but also for a subclass of renewal processes with varying variance (RPVV) as introduced in \citet{messer_paper}.

As the marginals of $L$ and $\widetilde{L}$ differ only in the $h$-neighborhood of $c$ and both processes are $2h$-dependent, their covariance structures differ in the $3h$-neighbourhood of the rate change point $c$. Our simulations in section \ref{subsec_signi_rinh} suggest that the differences between $L$ and $\widetilde{L}$ are typically  small  with respect to the $95\%$-quantile of their absolute maxima. We therefore suggest to use the parameter independent limit process $L$ also in the situation of potential rate change points for the derivation of the rejection threshold in the statistical test. The simulations in  section \ref{sec_eval} suggest that the MFT using $L$ instead of $\widetilde L$ keeps the asymptotic significance level for most combinations of $\mu$ and $\sigma$ even for the case of multiple unknown rate change points.

\section{Change point detection and  evaluation in simulations}
\label{sec_eval}

As the MFT is an asymptotic method, we investigate here the empirical significance level in simulations. Section \ref{subsec_signi_rconst} assumes a constant rate, and section \ref{subsec_signi_rinh} investigates the behavior for an unknown number of unknown change points in the rate. Section \ref{subsec_dp} evaluates the detection probability of variance change points in different simulation settings thereby showing the importance of including estimated rate change points and the dependence of the detection probability on the magnitude of changes and on the regularity of processes.

\subsection{Global rate} \label{subsec_signi_rconst}
\begin{figure}[ht!]
         	\centering
		\includegraphics[scale=0.48,clip=TRUE,trim= 0 0.0em 0 0.0em]{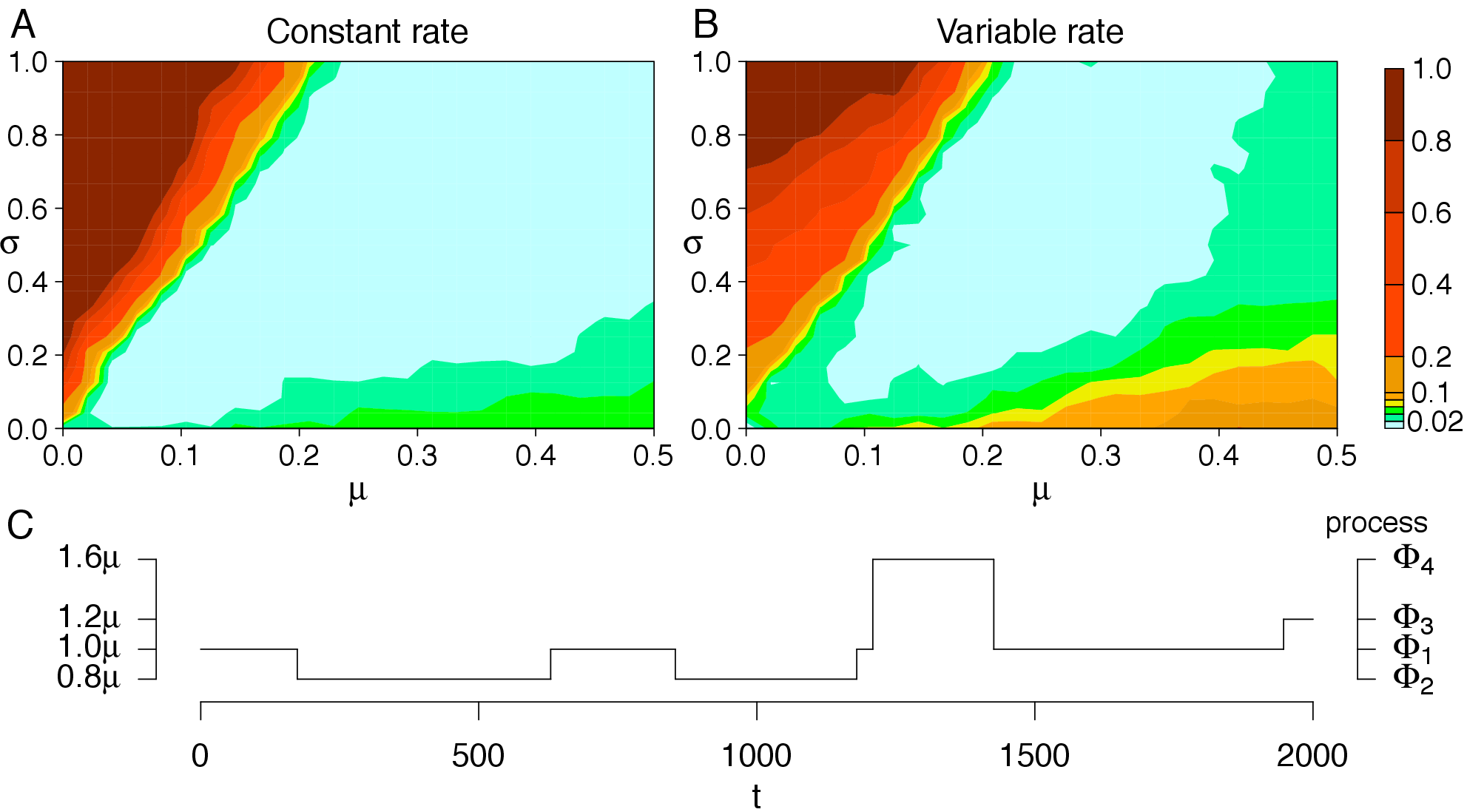}
         \caption{ Simulated rejection probability of the MFT for processes with i.i.d.~Gamma-distributed life times ($T=2000$, $H=\{150,250,500,750,1000,1250\} \cdot \mu$, 5000 simulations) (A) Constant unknown mean estimated by the global empirical mean. (B) The rate profile is given by a random change point model. For each simulation, a new rate profile is realized as exemplarily depicted in (C). The means and all change points are estimated using the MFA from \citet{messer_paper}. (C) Process $\Phi$ used in the simulations in (B) is a piecewise composition of four renewal processes $\Phi_1, \ldots, \Phi_4$ with Gamma-distributed life times with parameters  $(\mu_1, \mu_2, \mu_3, \mu_4) = (\mu, 0.8 \mu, 1.2 \mu, 1.6 \mu)$. Waiting times between  rate change points are uniformly distributed on $[0,800]$. At odd valued change points $\Phi$ jumps from  $\Phi_1$ to a randomly drawn other process, jumping back at even valued change points.}
					\label{fig_sig_level_mft}
\end{figure}

Figure \ref{fig_sig_level_mft} A shows the empirical significance level of the MFT applied to processes with independent and Gamma-distributed life times with mean $\mu$ and  standard deviation $\sigma$. The global empirical mean of the life times is used as an estimator for $\mu$. As discussed  in \citet{messer_paper}, the minimal window should be large enough such that a sufficiently high number of events can be observed. For change points in the rate, the minimal window should contain at least about 100-150 events \citep[see][]{messer_paper}. For variance change points, the minimal window should be slightly larger. We use here the window set $H=\{150,250,500,750,1000,1250\} \cdot \mu$, where the minimal window size increases linearly with the mean life time. As indicated in the figure, the test keeps the asymptotic significance level of $5\%$ for a wide range of parameter combinations and not too irregular processes (i.e., $\sigma<4\mu$). 

In the following section, the significance level of the MFT is investigated for a set of multiple unknown change points, implying also an unknown rate profile.

\subsection{Inhomogeneous rate}
\label{subsec_signi_rinh}
In order to investigate the significance level of the MFT in the case of multiple unknown rate change points, we first need to estimate the number and location of the rate change points. To this end, we apply here the multiple filter algorithm (MFA) for the rate proposed in \citet{messer_paper}. After estimation of the rate change points, we include the  estimated rates into the variance estimation in order to test the null hypothesis of variance homogeneity. When this null hypothesis is rejected, the MFA procedure can be extended to estimate the variance change points. In section \ref{sect:MFA1}, we  summarize the idea of the MFA and its two-step application for the detection of rate and variance change points. Section \ref{sect:MFA2} will  be concerned with  investigating its significance level in simulations with multiple change points in the rate.

\subsubsection{The two-step MFA for the detection of rate and variance change points} \label{sect:MFA1}
In a nutshell, the MFA works as follows. In case of rejection of the null hypothesis, change points are detected using the individual windows. For each window, we successively identify the maxima of the filtered derivative process and exclude their $h$-neighborhoods. In order to combine change points detected by different windows, change points are added  successively starting with the smallest windows. We then successively include only change points whose $h$-neighborhood does not contain already detected change points. For more details on the MFA compare \citet{messer_paper}. 

\begin{figure}[htbp]
          	\centering
          	\includegraphics[scale=0.46,clip=TRUE,trim= 0 0.0em 0 0.0em]{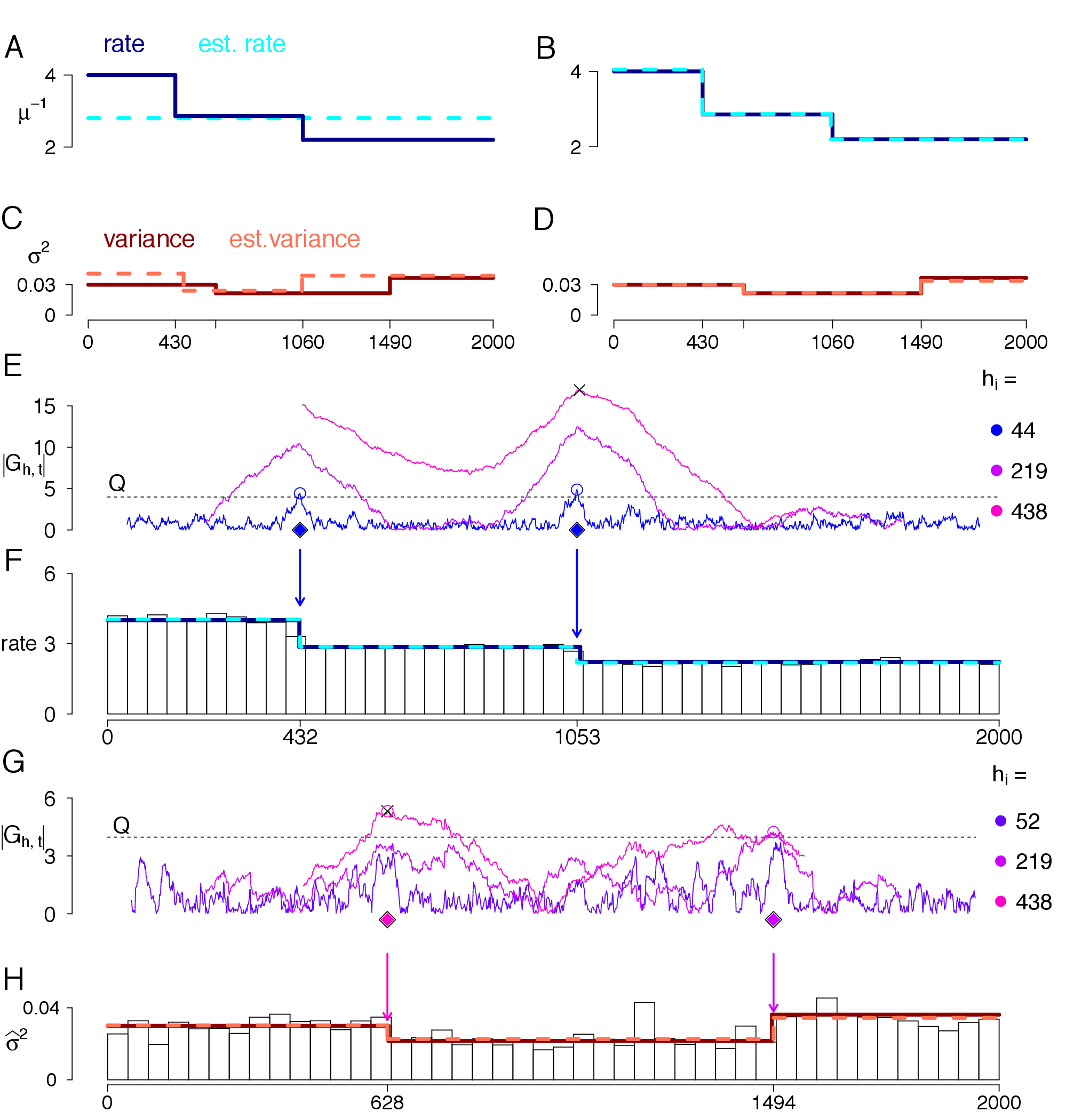}
\caption{Application of the sequential MFA for estimation of rate and variance change points in a simulated point process on $(0,2000]$ with rate change points $c_1=430, c_3=1060$ and variance change points $c_2=630, c_4=1490$. All life times were Gamma-distributed with $(\mu,\sigma^2)$ equalling $(0.25,0.03)$ in $(0,c_1]$,$(0.35,0.03)$ in $(c_1,c_2]$, $(0.35,0.0216)$ in $(c_2,c_3]$, $(0.45,0.0216)$ in $(c_3,c_4]$ and $(0.45,0.0357)$ in $(c_4,2000]$. (A, C) Neglecting rate inhomogeneity in the variance estimation yields erroneous results. Estimated rate (blue) and variance (red), and true profiles (darkblue, darkred). (B, D) The rate change points are estimated and included in the variance estimation. (E, F) Rate MFA. (G, H) Variance MFA. Colored curves show the $(|G_{h,t}|)$-processes colored by window size indicated on the right. Dashed line indicates simulated threshold $Q$, estimated change points are marked by diamonds. Dashed blue and red lines show the estimated rate and variance profiles, solid lines indicate the true parameter values.}
						\label{fig_application_simulated}
\end{figure}

We suggest to apply the MFA first for the estimation of rate change points. Second, we include the  estimated rates into the variance estimation (eq.~(\ref{eq_sigma2})). This sequential MFA  is illustrated in Figure \ref{fig_application_simulated} in a simulated point process with two rate and two variance change points. In panels B and D-H the estimated variance fits well if the inhomogeneous rates are included in the estimation. Panels A and C also indicate that neglecting the rate change points and thus estimating a constant rate results in erroneous estimation of the rate and the variance profile. This is because the applied test statistic uses the wrong global mean (eq.~(\ref{eq_def_vi_dach})).

\begin{figure}[ht!]
          	\centering
          	\includegraphics[scale=0.46,clip=TRUE,trim= 0 0.0em 0 0.0em]{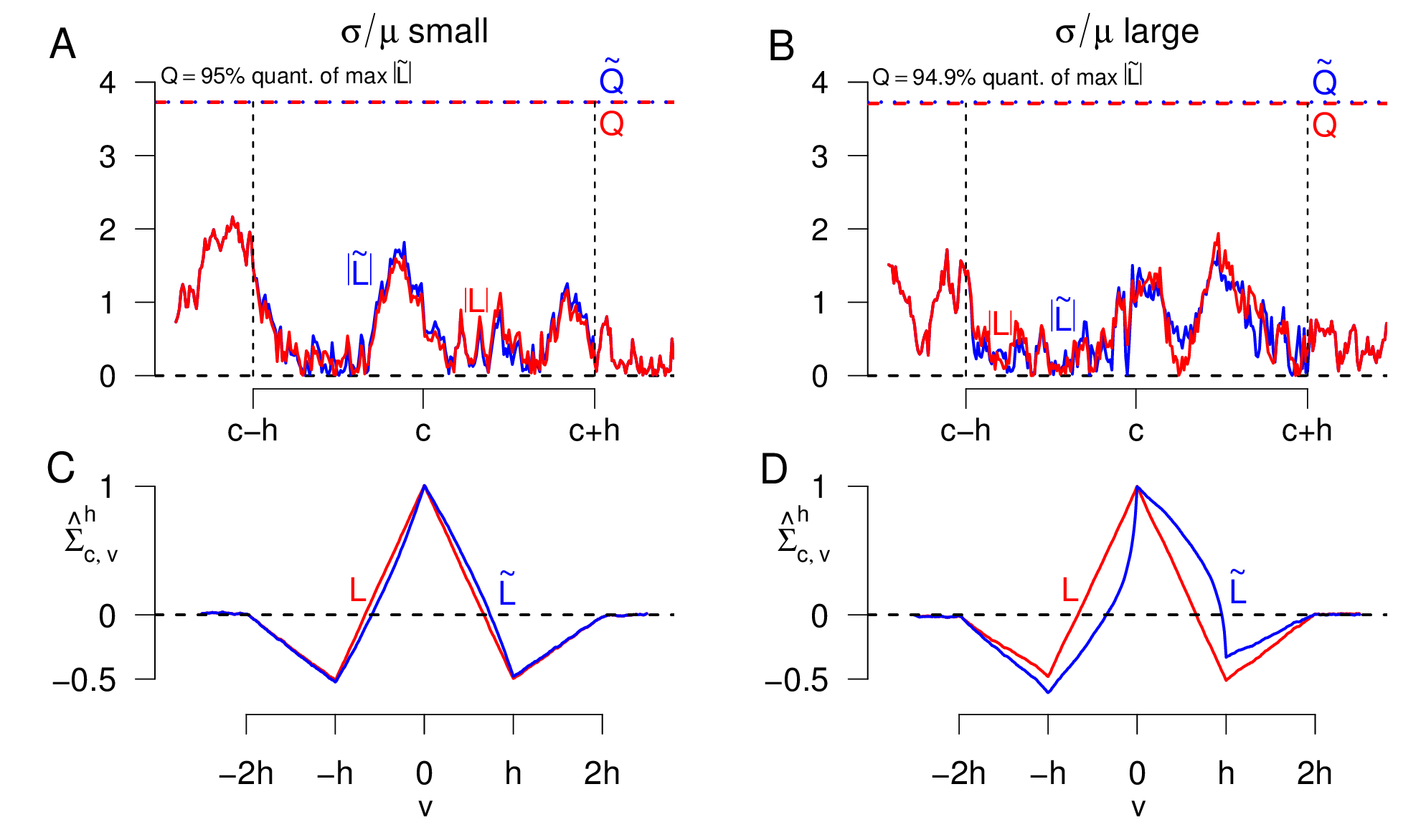}
         	\caption{Comparison of the processes $L$ and $\widetilde L$ in the neighborhood of a rate change point at $c=T/2$ ($T=2000$) for processes with small $\sigma/\mu$ and small rate change (A,C) and larger $\sigma/\mu$ and larger rate change (B,D).
	A, B: Realizations of $|L|$ (red) and $|\widetilde{L}|$ (blue), derived from the same Brownian motion. Outside the $h$-neighborhood of the change point, the marginals coincide. The $95\%$-quantiles $Q$ and $\widetilde{Q}$ of the absolute maxima of $L$ and $\widetilde L$ (estimated in 10000 simulations) are indicated by dashed and dotted lines. 
	C, D: The estimated empirical covariance functions $\hat \Sigma_{c,v}^h$ of $L$ (red) and $\widetilde {L}$  (blue) at the change point $c$ for one window size $h=100$ estimated in $10000$ simulations for the parameters given in A, B, respectively. Parameters for A, C: $\mu_1=0.1, \mu_2=0.15, \sigma_1=\sigma_2=0.1$. Parameters for B, D: $\mu_1=0.1, \mu_2=0.5, \sigma_1=\sigma_2=0.5$.} 
\label{fig_l_comp}
\end{figure}

Note that this procedure requires consistency of the estimated rate (Theorem \ref{main_theorem_b}). Although this has not been shown for the MFA, our simulation results suggest good performance (see section \ref{sect:MFA2}). In addition, note that in the second step of the sequential procedure, i.e., the detection of variance changes, the limit process $\widetilde L$ required to set the rejection threshold $Q$ differs from $L$. However, as $\widetilde L$ depends on  unknown process parameters, we argue here that one can replace $\widetilde L$ by $L$ because the mean and variance of  the two Gaussian processes are identical. Differences occur only in the covariance function $\Sigma_{u,v}^h:=\text{Cov}(L_{h,u},L_{h,u+v})$ in the $3h$-neighborhood of a change point and are typically small (Figure \ref{fig_l_comp} A, C), particularly for small $\sigma/\mu$ and small rate changes. For higher changes in the mean and higher $\sigma/\mu$, larger differences can be observed between $L$ and $\widetilde L$ (panel B, D), but their $95\%$-quantiles $Q$ and $\widetilde{Q}$ remain close together. Also in larger simulations with different rate changes up to a factor of six, $Q$ ranged between the $94.5\%$- and the $95.1\%$-quantile of  $\max_{h,t}|{\widetilde{L}}_{h,t}|$ (data not shown).

\subsubsection{Significance level under rate inhomogeneity} \label{sect:MFA2}
In order to investigate the empirical significance level of the MFT for variance changes  under an unknown set of unknown change points estimated with the rate MFA, we use a random rate change point model with rate changes of different height and at different time scales (Figure \ref{fig_sig_level_mft} C). The empirical significance level of the resulting MFT derived in 5000 simulations is plotted in Figure \ref{fig_sig_level_mft} B as a function of the mean $\mu_1$ of the process $\Phi_1$ and the standard deviation $\sigma$. 

Compared to the situation with constant rate (Figure \ref{fig_sig_level_mft} A), a higher type I error is observed as the rate MFA can usually not correctly estimate all rate change points, which affects the MFT for variance changes (Figure \ref{fig_application_simulated} C). The parameter region with empirical significance level $> 5\%$ is larger than under rate homogeneity, including also parameter combinations with high mean and small variance. This suggests that if the average variance is not too large or too small compared to the mean, the smallest window in the MFT for variance changes should contain at least about $150$  events.

\subsection{Detection probability of variance change points}
\label{subsec_dp}
\begin{figure}[h!]
         	\centering
		\includegraphics[scale=0.215,clip=TRUE,trim= 0 0.0em 0 0.0em]{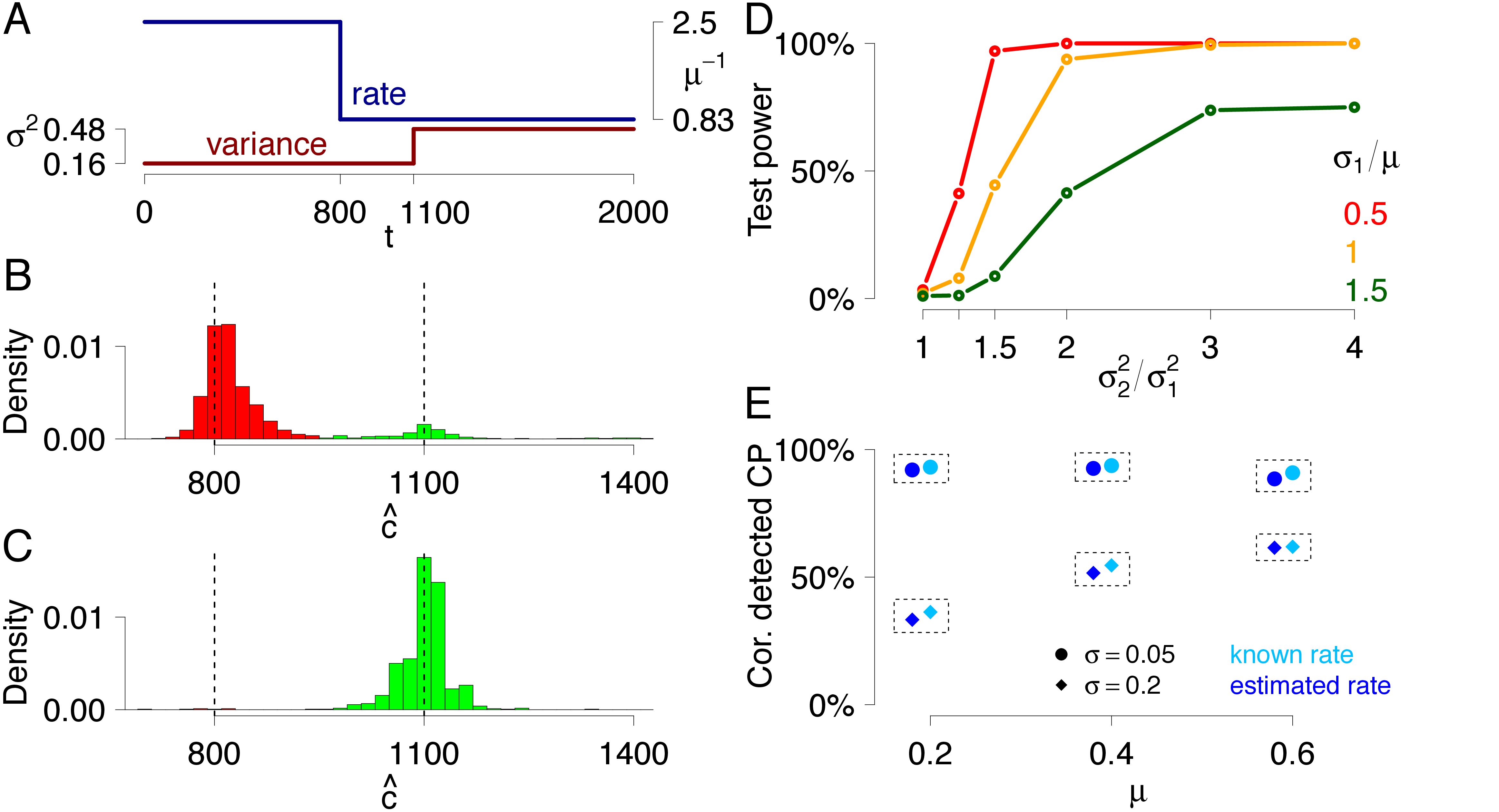}
         	\caption{Detection probability of variance change points, using $T=2000$, $\alpha=5\%$ and $H=\{150,250,500,750,1000,1250\} \cdot \mu$ throughout. A: Rate and variance profile of  the Gamma processes used in B and C. 
	B: Locations of estimated variance change points when the rate is assumed constant and estimated globally. For illustration, estimated change points closer to the (falsely detected) rate change point are colored in red, change point estimates closer to the variance change point are colored in green. C: Locations of estimated variance change points when the MFA for rate change detection is included as a first step. Colors as in B. D: Test power of the MFT for  variance homogeneity for Gamma processes with constant mean $\mu=0.4$ and one variance change point at $c=1000$ where the variance changes from $\sigma_1 \in \{0.2, 0.4, 0.6\}$ (cmp.~colors) to $\sigma_2^2$. E: Relative frequency of correctly detected variance change points in a random change point model for known inhomogeneous rates (light blue) and estimated inhomogeneous rates (dark blue). 
Rate and variance changes occur randomly with distances uniformly distributed on $[90,770]$.  At odd valued change points, the rate, the variance or both parameters change, each with probability one third. 
The variance changes uniformly from $\sigma^2$ to one of $(3 \sigma^2, 4 \sigma^2, 5\sigma^2)$, and the rate changes uniformly to one of $(0.4 \mu, 0.8 \mu, 1.2 \mu, 1.6 \mu)$, switching back to the original parameters $(\mu, \sigma^2)$ at even valued change points.  $5000$ simulations per data point. }
\label{fig_dp}
\end{figure}

Here we investigate the empirical detection probability of variance change points in simulations, considering cases with homogeneous and with inhomogeneous rate. First, we recommend to always perform the two-step procedure of estimating rate change points first and then using these for the analysis of variance homogeneity and estimation of variance change points. This is because in practice,  information about rate homogeneity is usually not given, and falsely assuming rate homogeneity can largely affect the analysis of variance homogeneity. As shown in Figure \ref{fig_dp} B, rate change points can be falsely identified as variance changes points, while the detection probability of true variance change points can dramatically decrease. 

Using this two-step procedure raises the question of whether the rate-MFT in the first step is applicable in the presence of variance change points. Indeed, one can show that the impact of variance change points on the performance of the rate-MFT is practically negligible \citep[for details see][Corollary 3.4]{messer_shark}. The reason is that if only the variance changes but not the rate,  the associated filtered derivative process for the rate still converges to a zero-mean unit-variance $2h$-dependent Gaussian process, and the change in the variance  affects only the local covariance structure of the limit process.  

Second, the MFT for variance changes shows a considerably high detection probability (Figure \ref{fig_dp} D). In accordance with common neurophysiological models, we simulated Gamma-distributed life times and call processes with life time distributions with a coefficient of variation (CV $=\sqrt{\mathbb Var(\xi)}/\mathbb E(\xi)$) of up to $0.5$ regular, while processes with CV$=1$ (e.g., a Poisson process) are called irregular, and processes with CV$>1$ very irregular. In regular and mildly irregular cases, a variance change factor of only $1.5$ already had a considerable detection probability of $50\%$ in the worst case, increasing quickly to detection probabilities close to $100\%$ for a change factor of $2$ \citep[cmp.~e.g.,][]{eckley2010, killick2010}. Only for extremely irregular cases, detection probability increased more slowly, reaching a detection probability of about 75\% at a change factor of $3$. 

Finally the proposed two-step procedure showed high performance in random change point models with multiple rate and variance changes. Figure \ref{fig_dp} E shows the percentage of correctly detected variance change points for different parameter combinations, where a  change point is called \textit{correctly detected} if it is contained in a neighborhood of at most $15$ time units of an estimated change point. 
Figure  \ref{fig_dp} E also shows that the detection probability of variance change points was not strongly affected by the necessity to estimate inhomogeneous rate profiles if estimated rate change points were included in the procedure. The percentage of correctly detected change points in simulations with unknown (dark blue) and known (light blue) inhomogeneous rate profiles were highly comparable. This is because rate changes that fail to be detected are typically too small to considerably affect the second step of estimating variance change points. All simulations were based on i.i.d.~Gamma-distributed life times, and similar results were also obtained with lognormally distributed life times (data not shown).

In summary, our simulations suggest good performance and practical applicability of the proposed two-step procedure of first detecting rate changes and then incorporating these estimates in the detection of variance change points. The significance level was kept for typical parameter constellations, and detection probability of variance change points was high even in the presence of multiple rate changes, as often observed in empirical data sets. This allows the analysis of empirical point processes with multiple rate and variance changes as illustrated in section \ref{sec_data}.

\section{Application to spike train recordings}
\label{sec_data}
In order to illustrate practical application of the proposed method, we analyze $72$ empirical spike train recordings of durations between  $540$ and $900$ seconds which were reported partly in \citet{schiemann2012} and analyzed for rate homogeneity in \citet{messer_paper}. As the mean firing rate was about $6$ Hz, the window set $H_R:=\{25,50,75,100,125,150\}$ was used there, and rate change points were estimated with the MFA. Here, we use these estimates of rate change points to analyze changes in the variance of  inter spike intervals. In order to ensure about $150$ events in the smallest window (see~section \ref{subsec_signi_rinh}), we chose a window set $H_V=H_R$. The significance level was set to $\alpha= 5\%$.

In $36$ out of all $72$ spike trains the null hypothesis of variance homogeneity was rejected, and in $22$ spike trains more than one variance change point was detected. In $11$ cases, different change points were detected by different window sizes. The mean rate of detected variance change points was about $0.1$ per minute. To measure the strength of a detected variance change we used the absolute difference of the estimated variances $|\hat\sigma_1^2 - \hat\sigma_2^2|$ normed with their mean $0.5(\hat\sigma_1^2+\hat\sigma_2^2)$. This strength ranged between $0.02 - 1.96$.  The detected variance ratios of changes ranged between $1.02-94\%$, where 53.6\% were below $2$ and even 82.9\% below $3$. Thus, a majority of detected variance changes showed variance ratios smaller than $3$ or even $2$, which indicates a high sensitivity of the proposed method also to comparably small variance changes.

Combined with the results of the rate change point detection, both null hypotheses of  rate and  variance homogeneity were rejected in about $50\%$ of all spike trains ($35$ out of $72$). For $27$ spike trains, only rate homogeneity was rejected, in one spike train only variance homogeneity was rejected, and for $9$ spike trains, neither null hypothesis was rejected.

\begin{figure}[htbp]
         	\centering					
		\includegraphics[scale=0.46,clip=TRUE,trim= 0 0.0em 0 0.0em]{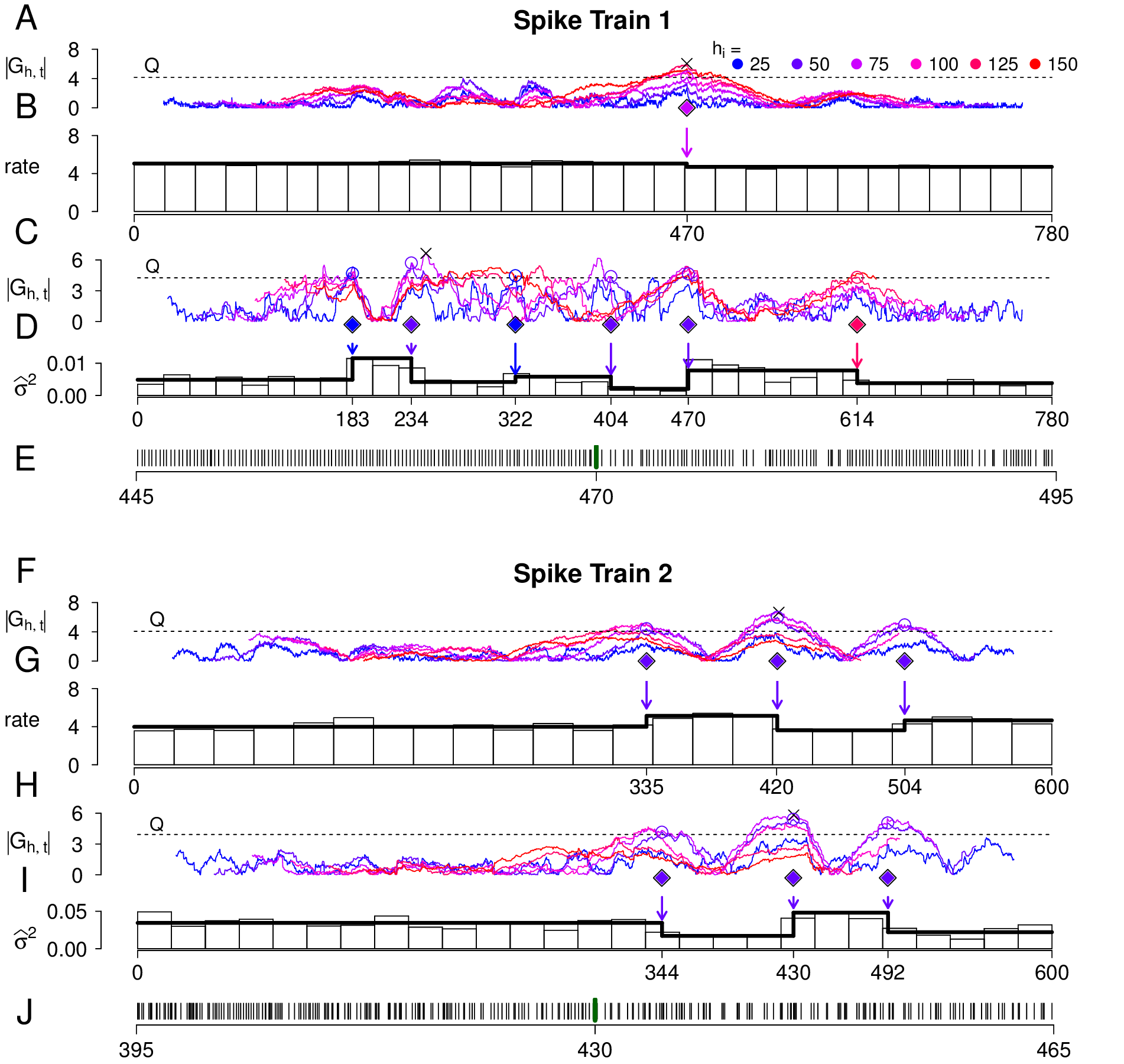}
         	\caption{Application of the rate and variance MFT to two spike train recordings; $H_{R}=H_{V}=\{25,50,75,100,125,150\}$, $\alpha=5\%$. (A), (C), (F) and (H) the processes $G$. The window sizes are color coded (legend on the upper right), and the  dashed line indicates the asymptotic rejection threshold $Q$. Diamonds mark the detected change points. In the first spike train (A-E) one rate and six variance change points are detected with four different windows. In the second spike train (F-J) one window detected three rate and three variance change points. (B) and (G) Rate histograms of the spike trains with estimated rates (black). (D) and (I) Variance histograms with estimated variances. (E) and (J) Parts of the spike trains with marked variance change points.}	
  \label{fig_data}			
\end{figure}

Figure \ref{fig_data} illustrates two spike train analyses with multiple rate and variance change points in which visual impression corresponds closely with the rate and variance profiles estimated by the algorithm. In the first example the rate only changes slightly while the variance shows six strong changes. In the second example, estimated rate and variance change points occur interestingly close to each other.

These findings stress that spike trains can show highly variable firing patterns, including a number of changes not only in the firing rate but also in the variability of inter spike intervals. Therefore, their detection prior to further analysis is strongly recommended when statistical analyses are sensitive to parameter changes.

\section{Summary and Discussion}
In this paper we have extended a multiple filter test (MFT) that has been proposed in \citet{messer_paper} and that aims at testing the null hypothesis of rate stationarity in renewal processes and to detect rate change points on multiple time scales. The rejection threshold of the test is derived from a Gaussian process $L$, which emerges as the limit of the filtered derivative process $G$ under stationarity and is independent of the point process parameters. 

By replacing the number of events in $G$ by the variance of life times, we analyze here homogeneity of the variance of the life times. In the presence of rate change points, the process $G$ deviates from zero in expectation in the neighborhood of the rate change point if $G$ is not adjusted for the rate change. This may lead to false interpretation of a rate change point as a variance change point. Therefore we propose an adaptation of the process $G$ that corrects for this deviation by taking into account the rate change. The resulting limit process $\widetilde L$ of $G$ vanishes in expectation in the neighborhood of a change point, but its covariance shows slightly different properties from $L$, and these differences depend on unknown process parameters. 

In practice, we propose to estimate the rate change points first by procedures that allow for potential variance changes \citep[e.g., the MFT for renewal processes with varying variance, ][]{messer_paper}. One can then incorporate these estimates in the statistical test for variance homogeneity. This is important in order to prevent false detection of rate change points as variance change points. As $\widetilde L$ depends on unknown process parameters, we use the process $L$ instead to compute asymptotic rejection thresholds. Our simulations suggest that the deviations between the limit processes tend to be small for a wide range of parameter values and that the use of $L$ does in these cases not considerably change the  properties of the statistical test. In simulations of point processes with constant variance and random rate change points the asymptotic significance level was kept if the smallest window was chosen sufficiently large. In addition, the simulations suggest that the detection of simulated variance change points is hardly affected by the necessity to estimate potential rate change points. 

In summary, we have extended a statistical test for the null hypothesis of rate homogeneity to the analysis of variance homogeneity in renewal processes with a wide range of life time distributions.  In addition, an algorithm is described that aims at detecting an unknown number of rate and variance change points that may occur at multiple time scales. When applying the procedure to empirical spike trains, both null hypotheses of constant rate and constant variance were rejected in the majority of cases, and multiple rate and variance change points were estimated. This suggests that the proposed method can be helpful for change point estimation and segmentation of empirical processes such as neuronal spike trains. It can thus be used as a means for signal detection or as a preprocessing step to statistical analyses that are sensitive to rate or variance changes.

\section*{Acknowledgements}
This work was supported by the German Federal Ministry of Education and Research (BMBF, Funding number: 01ZX1404B) and by the Priority Program 1665 of the DFG (SCHN 1370/2-1). 

\begin{appendix}
\section{Comparison of detection methods for change points in the mean}
\label{sect:comparison}
As mentioned in the introduction, the MFT is comparable to other change point methods. As an example, we show here a comparison to the penalized least squares (PLS) method proposed by \citet{lavielle2000} (Table \ref{tab:CompMFTPLS}). To allow for a comparison, the MFT is adjusted to sequences of random variables, and we test for changes in their mean. First (A), we simulated time series of $1000$ normally distributed random variables with variance $\sigma^2=1$ and $\mu_1=2, \mu_2=0, \mu_3=1$ with change points $c_1=250, c_2=500$.  This setting is comparable to the one in \citet{lavielle2000} but assumes independence according to the assumptions of the present MFT \citep[for an MFT extension to weak dependencies compare][]{messer2016}. Second (B), we applied a random change point model. We simulated $1000$ normally distributed random variables with mean $\mu_2=0$ and $\mu_1 \sim \mu_3 \sim \text{Unif}(0.4,2)$. The positions of the two change points were drawn uniformly without replacement out of $\{101, 102, \ldots, 900\}$ with a minimum distance of $100$ between the change points.  Both procedures yielded comparable mean absolute deviations of the estimated and the true change point locations (A: 2.5 (MFT) and 4.7 (PLS); B: 5.4 (MFT) and 5.7 (PLS).
\begin{table}[htbp]
	\centering
	\begin{tabular}{| l || l | l | l|}
			\hline
	 \multicolumn{4}{|c|}{\textbf{A} deterministic CPs  } 	\\
		\hline
		method    & 1 CP  & 2 CPs   & 3 CPs   \\
		\hline
		MFT & 0\%  & 99 \% & 1\% \\
		PLS & 0\%  & 100 \% & 0\% \\
		\hline
	\end{tabular}%
\hspace*{2em}
	\begin{tabular}{| l || l | l | l|}
		\hline
		\multicolumn{4}{|c|}{\textbf{B} random CPs  } 	\\
	\hline
	method    & 1 CP  & 2 CPs   & 3 CPs   \\
	\hline
	MFT & 3.6\%  & 95.4 \% & 1.0\% \\
	PLS & 3.0\%  & 94.9 \% & 2.1\% \\
	\hline
\end{tabular}%
	\captionof{table}{Comparison of the adjusted MFT and PLS  to detect changes in the mean. 1000 simulations with two CPs were performed for each scenario, and the percentage of simulations with one, two or three detected CPs are indicated. Throughout the simulations, the adjusted MFT used the window set $H=\{100,200,300,400\}$. For PLS we used the penalty terms $\beta_n=\log(1000)$ in A and $\beta_n=0.5\log(1000)$ in B.}
	\label{tab:CompMFTPLS}
\end{table}%
\section{Proofs of Theorems}

The proof of Theorem \ref{main_theorem_a} in section \ref{sec_proof_a} uses the Anscombe-Donsker-Theorem and the consistency of the estimator $\hat{s}^2$. In section \ref{sec_proof_b}, we prove Theorem \ref{main_theorem_b} basically using the same ideas as for the proof of Theorem \ref{main_theorem_a}. 

Throughout the Appendix we use the following notation. For $\tau > 0$ we denote the set of all c\`{a}dl\`{a}g functions on $[0,\tau]$ by $D[0,\tau]$. $d_{||\cdot||}$ serves as abbreviation for the metric induced by the supremum norm.  The Skorokhod metric on $D[0,\tau]$ is abbreviated by $d_{SK}$.  We use $D[0,\infty)$ and $D[\tau_h]$ with the Skorokhod metric. Note the fact that convergence in $(D[0,\infty),d_{||\cdot||})$ implies convergence in $(D[0,\infty),d_{SK})$. Furthermore, for an a.s. constant stochastic process in $D[\tau_h]$ with value $c$ we abbreviate the process $(c)_{t \in \tau_h}$ with $c$. Note that uniform a.s. convergence interchanges with sums in general and with products if the limits are constant.  
For a point process $\Phi$ with events $(S_i)_{i \geq 1}$ the corresponding counting process $(N_t)_{t \geq 0}$ is defined as
\begin{align*}
N_t:= \max \{i \geq 1 | S_i \leq t\}, \quad t \geq 0, 
\end{align*}
with the convention $\max \emptyset := 0$. We specify the sets of indices
\begin{align*}
\hat I_{\text{le}}:=\{N_{n(t-h)}+2, N_{n(t-h)}+3, \ldots, N_{nt}\} 
\quad \quad \text{and} \quad \quad
\hat I_{\text{ri}}:=\{N_{nt}+2, N_{nt}+3, \ldots, N_{n(t+h)}\} .
\end{align*}

\subsection{Proof of Theorem \ref{main_theorem_a}}
\label{sec_proof_a}
The main ingredients for the proof of weak convergence of the filtered derivative process
\begin{align*}
G_t^{(n)}=\frac{1}{\hat s_t^{(n)}} \left(\frac{1}{N_{n(t+h)}-N_{nt}-1}\sum\limits_{i=N_{nt}+2}^{N_{n(t+h)}}{\hat V_i} - \frac{1}{N_{nt}-N_{n(t-h)}-1}\sum\limits_{i=N_{n(t-h)}+2}^{N_{nt}}{\hat V_i}\right)
\end{align*}
 in Theorem \ref{main_theorem_b} are the Donsker-Anscombe-Theorem and continuous mapping. In step 1 we assume a known mean $\mu$ and a known $s$ and thus use an auxiliary process  $\Gamma^{(n)}$ defined as follows
 \begin{align*}
\Gamma_t^{(n)}:= \Gamma_{\text{ri},t}^{(n)}-\Gamma_{\text{le},t}^{(n)}:= \frac{1}{s_t^{(n)}} \left(\frac{1}{N_{n(t+h)}-N_{nt}-1}\sum\limits_{i=N_{nt}+2}^{N_{n(t+h)}}{V_i} - \frac{1}{N_{nt}-N_{n(t-h)}-1}\sum\limits_{i=N_{n(t-h)}+2}^{N_{nt}}{V_i}\right). 
\end{align*}
We show that in $(D[h,T-h] \times D[h,T-h], d_{SK} \otimes d_{SK})$  it holds as $n \to \infty$
\begin{align}
\label{eq:joint_cov}
\left(\left(\Gamma_{\text{ri},t}^{(n)}\right)_{t \in \tau_h}, \left(\Gamma_{\text{le},t}^{(n)} \right)_{t \in \tau_h} \right) \stackrel{d}{\longrightarrow} \left(  \left(\frac{W_{t+h}-W_t}{\sqrt{2h}}\right)_{t \in \tau_h}  ,  \left(\frac{W_{t}-W_{t-h}}{\sqrt{2h}}\right)_{t \in \tau_h}  \right)
\end{align}
which yields $\Gamma^{(n)} \xrightarrow{d} L$.  \\
In step 2 the true mean $\mu$ is replaced by the estimated global mean $\hat \mu$ and $s$ is replaced by $\hat s$ thereby showing that $G^{(n)} \xrightarrow{d} L$ holds true. \\

\noindent \underline{Step 1: weak process convergence for known parameters}\\
Recall $V_i=(\xi_i-\mu)^2$ for the life times $(\xi_i)_{i \geq 1}$ of a point process $\Phi$ with a known mean $\mu$. We apply a FCLT for randomly stopped processes (Anscombe-Donsker version) to the process 
defined by
\begin{align}
Y_t^{(n)}:=\frac{1}{\nu \sqrt{n}} \sum\limits_{i=1}^{N_{nt}}{(V_i -\mathbb{E}[V_1])}.
\end{align}
Since $N_t/t\to 1/\mu$ a.s. as $n\to\infty$, it follows that in 
$(D[0,\infty),d_{SK})$ it holds $(\sqrt{\mu}Y_t^{(n)})_t\xrightarrow{d} (W_t)_t$, see e.g, \citet{gut}. Here, $W$ denotes a standard Brownian motion.
Let $\varphi: (D[0,\infty),d_{SK}) \to (D[h,T-h] \times D[h,T-h], d_{SK} \otimes d_{SK})$ be defined by
\begin{align*}
	(f(t))_{t \geq 0}	&\xmapsto{\varphi} \left(\left(\frac{(f(t+h)-f(t))}{\sqrt{2h}}\right)_{t \in \tau_h},\left(\frac{(f(t)-f(t-h))}{\sqrt{2h}}\right)_{t \in \tau_h}\right).
\end{align*}
This function is continuous. Mapping $\sqrt{\mu} Y^{(n)}$ via $\varphi$, the first component is given by 
\begin{align}\label{phi1}
\left(\sqrt{\frac{\mu}{2nh\nu^2}}\left[ \left(\sum_{i=N_{nt}+1}^{N_{n(t+h)}} V_i\right) - (N_{n(t+h)}-N_{nt})\mathbb{E}[V_1]\right]\right)_{t\in\tau_h} \stackrel{d}{\longrightarrow} \left(\frac{W_{t+h}-W_t}{\sqrt{2h}}\right)_{t \in \tau_h}.
\end{align}
 Now in $(D[h,T-h],d_{SK})$ it holds almost surely that $(N_{n(t+h)}-N_{nt}-1)_t \sim (nh/\mu)_t$ as $n\to\infty$,
see Lemma A.3.2 in \citet{messer_paper}. Thus by Slutsky's theorem, 
\begin{align*}
(\Gamma_{\text{ri},t}^{(n)})_{t \in \tau_h} \xlongrightarrow{d} \left(\frac{W_{t+h}-W_t}{\sqrt{2h}}\right)_{t \in \tau_h}
\end{align*}
(here, we also omitted one summand $i=N_{nt}+1$, i.e., a term of order $o_{a.s.}(1)$). By exchanging $t$ with $t-h$ and $t+h$ with $t$, we obtain the convergence for $\Gamma_{\text{le}}^{(n)}$ , which refers to the second component of $\varphi(\sqrt{\mu}Y^{(n)})$. Thus, we obtain (\ref{eq:joint_cov}).
This implies $\Gamma^{(n)}\stackrel{d}{\longrightarrow} L$ in $(D[h,T-h],d_{SK})$ as $n\to\infty$ by continuous mapping which is the assertion for a known mean $\mu$. Note that the expectation $\mathbb{E}[V_1]$ vanishes, since it appears in both summands.  \\

\noindent \underline{Step 2: replacement of parameters by their estimators} \\
In a second step we use the estimated mean $\hat \mu_{nT}$. We show that for 
\begin{align*}
	\hat Y_t^{(n)}:=\frac{1}{\nu \sqrt{n}} \sum\limits_{i=1}^{N_{nt}}{(\hat V_i - \mathbb{E}[ V_1])}
\end{align*}
in $(D[0,\infty),d_{SK})$ we also obtain
\begin{align}
\label{conv_ynttilde}
(\sqrt{\mu} \hat Y_t^{(n)})_{t \in [0,T]} \xlongrightarrow{d} (W_t)_{t \in [0,T]}.
\end{align}
Thus, the same arguments as in step 1 can be applied to show the assertion $G^{(n)} \xrightarrow{} L$. Additionally, by Slutsky's theorem and Corollary \ref{corr_s_kons} (given below), we can exchange the factor $(nh/2\mu\nu^2)^{1/2}$ with the estimator $1/\hat s(nt,nh)$ with the convergence holding true. 

\noindent To show (\ref{conv_ynttilde}), we rewrite $\hat Y_t^{(n)}=  Y_t^{(n)} + R_t^{(n)}$, with $R$ given by
\begin{align*}
R_t^{(n)} := \frac{N_{nt}}{\nu \sqrt{n}}\left[ (\hat\mu_{nT}-\hat\mu_{nt})^2 - (\hat\mu_{nt}-\mu)^2 \right].
\end{align*}
This decomposition holds true since
\begin{align*}
\sum_{i=1}^{N_{nt}} \hat V_i
\;\,=\;\,
\sum_{i=1}^{N_{nt}} (\xi_i-\hat\mu_{nT})^2 
&= 
\sum_{i=1}^{N_{nt}} \xi_i^2 - 
2\hat\mu_{nT} \sum_{i=1}^{N_{nt}}\xi_i
+N_{nT} \hat\mu_{nT}^2\\
&=
\sum_{i=1}^{N_{nt}}(\xi_i-\mu)^2
+ 2(\mu-\hat\mu_{nT})\sum_{i=1}^{N_{nt}}\xi_i
+ N_{nt}(\hat\mu_{nT}^2-\mu^2)\\
&=
\sum_{i=1}^{N_{nt}} V_i
+ N_{nt} \left[ (\hat\mu_{nT}-\hat\mu_{nt})^2 
 - (\hat\mu_{nt}-\mu)^2\right].
\end{align*}
 We now show that in $(D[0,T],d_{||\cdot||})$ the remainder $R \to 0$ in probability as $n\to\infty$. Then, the convergence in (\ref{conv_ynttilde}) follows by Slutsky's theorem. \\
 It suffices to show that  $\sup_{t\in[0,T]} R_t^{(n)}$ vanishes in probability. Recall $\sigma^2=\mathbb{V}\!ar(\xi_1)$ and set
\begin{align*}
Z_t^{(n)}:= \frac{\sqrt{\mu}}{\sigma \sqrt{n}}\sum_{i=1}^{N_{nt}}(\xi_i-\mathbb{E}[\xi_i]) = \frac{\sqrt{\mu} N_{nt}}{\sigma\sqrt{n}}(\hat\mu_{nt}-\mu).
\end{align*}
Then, by Donsker-Anscombe theorem we find in $(D[0,T],d_{SK})$ that $(Z_t^{(n)})_{t\in[0,T]}\to (W_t)_{t\in[0,T]}$ weakly as $n\to\infty$, such that 
$\sup_{t\in[0,T]} Z_t^{(n)} \to \sup_{t\in[0,T]} W_t$ weakly. Now, we first focus on the second summand of $R_t^{(n)}$ and show that its square root vanishes
\begin{align*}
\sup_{t\in[0,T]}\left[\left(\frac{N_{nt}}{\nu \sqrt{n}}\right)^{1/2} (\hat\mu_{nt}-\mu) 
 \right]
=
\sup_{t\in[0,T]}\left[ \left(\frac{n \sigma^4}{(\mu \nu N_{nt})^{2}}\right)^{1/4} Z_t^{(n)}
 \right]
 \stackrel{\mathbb{P}}{\longrightarrow} 0,
\end{align*}
as $n\to\infty$. This holds true since in $(D[0,T],d_{SK})$ it holds almost surely that $(N_{nt})_t \sim (nt/\mu)_t$ as $n\to\infty$, compare e.g., Lemma A.3.2 in \citet{messer_paper}. Thus, the first factor within the supreme itself vanishes such that the entire expression tends to zero. By continuous mapping theorem, this holds true for the squared expression, such that the second summand in $R_t^{(n)}$ uniformly tends to zero in probability. For the first summand in $R_t^{(n)}$, we decompose  $(\hat\mu_{nT}-\hat\mu_{nt}) = (\hat\mu_{nT} - \mu ) -(\hat\mu_{nt}-\mu)$ and apply the same argument as before to both summands. As a result 
$R$ uniformly vanishes in probability.  $\Box$

Now, we state the consistency of the variance estimator $\hat s$ in Corollary \ref {corr_s_kons}.
\begin{lemma}
\label{nu_kons}
Let $\Phi$ be an element of $\mathscr{R}$ with $\nu^2=\V((\xi_1-\mu)^2)$. Let $T>0$, $h \in (0,T/2]$ and $\hat{\nu}^2_\text{ri}$ and $\hat{\nu}^2_\text{le}$ be defined as in (\ref{def_nu_exact}) using the estimated global mean $\hat \mu^{(i)}=\hat \mu_{nT}=(1/N_{nT}) \sum_{i=1}^{N_{nT}} \xi_i$. Then it holds in $(D[h,T-h],d_{||\cdot||}))$ almost surely as $n \to \infty$ that
\begin{equation}\label{cons_nu} 
	(\hat{\nu}^2_\textrm{ri})_{t \in \tau_h} \longrightarrow (\nu^2)_{t \in \tau_h} \qquad \text{ and }\qquad (\hat{\nu}^2_\textrm{le})_{t \in \tau_h} \longrightarrow (\nu^2)_{t \in \tau_h}.
\end{equation}
\end{lemma}

\textit{Proof:} For a known mean $\mu$, i.e., $V_i$ instead of $\hat V_i$ convergences (\ref{cons_nu}) are shown using the same techniques as for the consistencies of $(\hat \mu)_t$ and $(\hat \sigma^2)_t$ in \citet{messer_paper}. A complete proof can be found in \citet{albert}.
 The extension to the estimated global mean $\hat \mu_{nT}$ follows from the strong consistency $\hat \mu_{nT}\to\mu$ a.s. as $n\to\infty$ (see e.g., \citet{messer_paper}) and by standard application of Slutsky's theorem. $\Box$\\

Lemma \ref{nu_kons} directly implies the consistency of the variance estimator $\hat{s}^2$.

\begin{corollary}
	\label{corr_s_kons}
	Let $\Phi \in \mathscr{R}$ with $\nu^2=\V((\xi_1-\mu)^2)$. Let $T>0$, $h \in (0,T/2]$ and $\hat{s}^2(t,h)$ be defined as in (\ref{def_s}). Then it holds in $(D[\tau_h],d_{||\cdot||})$ almost surely as $n \to \infty$ that
	\begin{equation*} 
	\left(n\;\hat{s}^2(nt,nh)\right)_{t \in \tau_h} \longrightarrow \left(\frac{2\nu^2}{h/ \mu}\right)_{t \in \tau_h}
	\end{equation*}
\end{corollary}

\textit{Proof:} This follows from Lemma \ref{nu_kons} by  application of Slutsky's theorem. $\Box$


\subsection{Proof of Theorem \ref{main_theorem_b}}
\label{sec_proof_b}
For the proof of weak convergence of the filtered derivative process in Theorem \ref{main_theorem_b}, we use again the Donsker-Ascombe-Theorem and continuous mapping. In addition to the previous proof, a change point in the rate requires separate considerations for different intervals in the neighborhood of a change point. These are different for the right and left window and therefore, we define auxiliary processes that correspond to the right and left window, respectively.

Like in the proof of Theorem \ref{main_theorem_a} we first assume known process parameters and use the modified filtered derivative process $\Gamma$ 
\begin{equation}
\Gamma_t^{(n)} =  \Gamma_{\text{ri},t}^{(n)}-\Gamma_{\text{le},t}^{(n)}.
\end{equation}
The latter terms are given by $\Gamma_{\text{ri}}^{(n)}=$
\begin{align}
		\label{def_gamma1}
		\begin{cases}
		\frac{\frac{1}{N_{n(t+h)}-N_{nt}-1}\left(\sum\limits_{i=N_{nt}+2}^{N_{n(t+h)}}{(\xi_i-\mu_1)^2}\right)-\sigma^2}{s^{(n)}_t}, &  \text{if } t < c-h, \\
		\frac{\frac{1}{N_{n(t+h)}-N_{nt}-1}\left(\sum\limits_{i=N_{nt}+2}^{N_{nc}}{(\xi_i-\mu_1)^2}+\sum\limits_{i=N_{nc}+2}^{N_{n(t+h)}}{(\xi_i-\mu_2)^2}\right)-\sigma^2}{s^{(n)}_t }, &  \text{if } c-h \leq t <c,\\
		\frac{\frac{1}{N_{n(t+h)}-N_{nt}-1}\left(\sum\limits_{i=N_{nt}+2}^{N_{n(t+h)}}{(\xi_i-\mu_2)^2}\right)-\sigma^2}{s^{(n)}_t }, &  \text{if } t \geq c,\\
		\end{cases} 
\end{align}
and analogously for $\Gamma^{(n)}_{\text{le}}$. Analogously, we decompose the limit process $\widetilde L_{\text{ri}}-\widetilde L_{\text{le}} \sim \widetilde L$, where $\sim$ denotes equality in distribution. With $(W_{1,t})_{t \geq 0}$ and $(W_{2,t})_{t \geq 0}$  independent standard Brownian motions the latter terms are given by $\widetilde{L}_{\text{ri}}:=\widetilde{L}_{\text{ri},h,t}=$
\begin{align}
\label{def_l1}
     	\begin{cases}
		\frac{(W_{1,t+h}-W_{1,t})}{\sqrt{2h}}, & \text{if } t < c-h, \\
			 \frac{\sqrt{\mu_{\text{ri},t}^2\nu^2_2 / (h^2 \mu_2)}(W_{2,t+h}-W_{2,c})+\sqrt{\mu_{\text{ri},t}^2 \nu^2_1 / (h^2 \mu_1)}(W_{1,c}-W_{1,t})}{s^{(1)}_t },	& \text{if } c -h\leq t <c,\\
			  \frac{\sqrt{\mu_2\nu^2_2 / h^2}(W_{2,t+h}-W_{2,t})}{s^{(1)}_t}, &   \text{if } t \geq c,\\	
					\end{cases}
		\end{align}
and analogously for $\widetilde L_{\text{le}}$.  The first step of the proof (see step $1$ below) will be to show convergence of the processes
\begin{equation}\label{eq:widetildegamma}
\widetilde \Gamma_{\text{ri}}^{(n)}:=\left(\frac{N_{n(t+h)}-N_{nt}-1}{nh/\mu_{\text{ri}}}\right)_t\cdot \Gamma_{\text{ri}}^{(n)}
\end{equation}
and $\widetilde \Gamma_{\text{le}}$ against $\left(\widetilde L_{\text{ri}},\widetilde L_{\text{le}}\right)$ using the Donsker-Ascombe-Theorem and continuous mapping. With Lemma \ref{lemma_nhprocess}, we can then conclude 
\begin{equation}
\left(\Gamma_{\text{ri}}^{(n)},\Gamma_{\text{le}}^{(n)}\right) \xlongrightarrow[]{d}  \left(\widetilde L_{\text{ri}},\widetilde L_{\text{le}} \right),
\end{equation} 
and using continuous mapping again yields for $t \in \tau_h$
\begin{equation}
\Gamma^{(n)}=\Gamma_{\text{ri}}^{(n)}-\Gamma_{\text{le}}^{(n)} \xlongrightarrow[]{d} \widetilde L_{\text{ri}}-\widetilde L_{\text{le}} \sim \widetilde L.
\label{eq:GammaToL}
\end{equation}
In step two of the proof, we first replace the true means $\mu_1, \mu_2$ in the numerator by their estimators to define the process $\hat{\Gamma}^{(n)}$ and show
\begin{equation}
\label{eq:hatGamma}
\hat{\Gamma}^{(n)}-\widetilde{\Gamma}^{(n)}  \xlongrightarrow[]{\mathbb{P}} (0)_t.
\end{equation}
Then, we use Lemma \ref{lemma_conv_s2} to substitute the scaling $s_t^{(1)}$ used in $\hat\Gamma$ by the estimator $\hat s_t^{(n)}$ in order to prove the assertion. \\

\noindent \underline{Step $1$:} Proof of \\
\begin{equation}
\left(\widetilde\Gamma_{\text{ri}}^{(n)},\widetilde\Gamma_{\text{le}}^{(n)}\right) \xlongrightarrow[]{d}  \left(\widetilde L_{\text{ri}},\widetilde L_{\text{le}} \right).
\end{equation} 
Let $(\xi_{1,i})_{i \geq 1}$, $(\xi_{2,i})_{i \geq 1}$ and $(\xi_i)_{i \geq 1}$ denote the sequences of life times that correspond to $\Phi_1$, $\Phi_2$  and to the compound process $\Phi$, respectively. Analogously, let $(N_{1,t})_{t \geq 0}$, $(N_{2,t})_{t \geq 0}$ and $(N_{t})_{t\geq0}$ denote the counting processes that correspond to $\Phi_1$, $\Phi_2$ and to $\Phi$. 
We use the abbreviated notation 
\begin{equation*}
V_{j,i}:=(\xi_{j,i}-\mu_j)^2
\end{equation*}
for the individual processes $\Phi_j$, $j=1,2$. According to the Anscombe-Donsker-Theorem we observe in $(D[0,\infty),d_{SK})$ as $n \to \infty$
\begin{align*}
\left(Z^{(n)}_{j,t}\right)_{t \geq 0}&:=\left(\frac{1}{\nu_j\sqrt{\frac{n}{\mu_j}}} \left( \sum\limits_{i=1}^{N_{j,nt}}{(V_{j,i}-\mathbb{E}[V_{j,i}])} \right)\right)_{t \geq 0}
\xlongrightarrow[]{d} (W_{j,t})_{t \geq 0}.
\end{align*}
Using a different scaling it holds in $(D[0,\infty),d_{SK})$ as $n \to \infty$
\begin{equation}
\label{eq:donsker}
\left(\widetilde{Z}^{(n)}_{j,t}\right)_{t \geq 0}:=\left(\sqrt{\frac{\nu^2_j}{nh^2 \mu_j}}\frac{1}{s^{(n)}_t}{Z^{(n)}_{j,t}}\right)_{t \geq 0} \xlongrightarrow[]{d} \left(\sqrt{\frac{\nu^2_j}{h^2  \mu_j}}\frac{1}{s^{(1)}_t }W_{j,t}\right)_{t \geq 0}
\end{equation}
because $(\sqrt{\nu^2_j /( nh^2  \mu_j)}/s^{(n)}_t))_t$ is continuous in $t$ and does not depend on $n$. As $\Phi_1$ and $\Phi_2$ are independent, we also obtain joint convergence  in $(D[0,\infty)\times D[0,\infty), d_{SK} \otimes d_{SK})$ as $n \to \infty$
\begin{equation}
\label{jointconv}
\left(\left(\widetilde{Z}^{(n)}_{1,t}\right)_{t \geq 0},\left(\widetilde{Z}^{(n)}_{2,t}\right)_{t \geq 0}\right) \xlongrightarrow[]{d} \left(\left(\sqrt{\frac{\nu^2_1}{h^2 \mu_1}}\frac{1}{s^{(1)}_t }W_{1,t}\right)_{t \geq 0}, \left(\sqrt{\frac{\nu^2_2}{h^2 \mu_2}}\frac{1}{s^{(1)}_t }W_{2,t}\right)_{t \geq 0}\right).
\end{equation}
For $ \mu_{\text{ri}}(t)= \mu_{\text{ri},t}$,  $\mu_{\text{le}}(t)= \mu_{\text{le},t}$ (as in eq. (\ref{def_muaverage2})), we use the map $\varphi: (D[0,\infty)\times D[0,\infty), d_{SK} \otimes d_{SK}) \xrightarrow []{}  (D[\tau_h] \times D[\tau_h],d_{SK}\otimes d_{SK})$ given by
\begin{align*}
	((f(t))_{t \geq 0},(g(t))_{t \geq 0})\xmapsto[]{\varphi} &		
 \left(	\left(
 	\begin{array}{ll} 
 (f(t+h)-f(t)) \mu_{\text{ri}}(t)\mathds{1}_{[h,c-h)}(t) \\
+(g(t+h)-g(c)) + (f(c)-f(t)) \mu_{\text{ri}}(t)\mathds{1}_{[c-h,c)}(t) \\
+(g(t+h)-g(t)) \mu_{\text{ri}}(t)\mathds{1}_{[c,T-h)}(t) \\
\end{array}
\right)_{t} , \right.\\
& \quad\quad\quad \left.\vphantom{ \left(
 \begin{array}{ll}
  (f(t)-f(t-h))\mu_{\text{le}}(t)\mathds{1}_{[h,c)}(t) \\
+(g(t)-g(c))+(f(c)-f(t-h)) \mu_{\text{le}}(t) \mathds{1}_{[c,c+h)}(t) \\
+(g(t)-g(t-h)) \mu_{\text{le}}(t) \mathds{1}_{[c+h,T-h]}(t) \\
\end{array}
\right)_{t}}  \left(
 \begin{array}{ll}
  (f(t)-f(t-h))\mu_{\text{le}}(t)\mathds{1}_{[h,c)}(t) \\
+(g(t)-g(c))+(f(c)-f(t-h)) \mu_{\text{le}}(t) \mathds{1}_{[c,c+h)}(t) \\
+(g(t)-g(t-h)) \mu_{\text{le}}(t) \mathds{1}_{[c+h,T-h]}(t) \\
\end{array}
\right)_{t} \right)
\end{align*}
\noindent As both component functions are compositions of continuous functions $\varphi$ is also continuous. The Continuous-Mapping-Theorem explains why convergence (\ref{jointconv}) holds with map $\varphi$ applied to both sides. 
$\varphi$ applied to the right hand side of (\ref{jointconv}) equals $\left(\widetilde L_{\text{ri}}(t), \widetilde L_{\text{le}}(t)\right)$ in distribution, which can be obtained by elementary calculations. For the left hand side of (\ref{jointconv}) we show
\begin{align}
\label{toshow_gamma}
\varphi\left(\left(\widetilde{Z}^{(n)}_{1,t}\right)_{t \geq 0},\left(\widetilde{Z}^{(n)}_{2,t}\right)_{t \geq 0}\right)  &= \left(\left(\widetilde \Gamma^{(n)}_{\text{ri},t}\right)_{t \in \tau_h}, \left(\widetilde \Gamma^{(n)}_{\text{le},t}\right)_{t \in \tau_h}  \right).
\end{align}
We make the first coordinate explicit. There, we distinguish between the three cases $t \in [h,c-h)$, $t \in [c-h,c)$ and $t \in [c,T-h]$.
For $t < c-h$, the first coordinate of the right hand side in equation (\ref{toshow_gamma}) is given by
\begin{align*}
\frac{1}{nh/ \mu_{\text{ri}}} \sum\limits_{i=N_{1,nt}+2}^{N_{1,n(t+h)}}{(V_{1,i}-\E[V_{1,i}])}\frac{1}{s^{(n)}_t  } 
&=\left(\frac{1}{nh/ \mu_{\text{ri}}}  \sum\limits_{i=N_{nt}+2}^{N_{n(t+h)}}{V_{i}}-\frac{N_{n(t+h)}-N_{nt}-1}{nh / \mu_{\text{ri}}}\E[V_i]\right)\frac{1}{s^{(n)}_t }.
\end{align*}
Exchanging subscripts yields analogous results for $t \geq c$. For $t \in [c-h,c)$ we obtain the first coordinate as 
\begin{align*}
&\frac{1}{nh/ \mu_{\text{ri}}} \left(\sum\limits_{i=N_{1,nt}+2}^{N_{1,nc}}{V_{1,i}-\E[V_{1,i}]}+\sum\limits_{i=N_{2,nc}+2}^{N_{2,n(t+h)}}{V_{2,i}-\E[V_{2,i}]}\right)\frac{1}{s^{(n)}_t} \\
&=\left(\frac{1}{nh/ \mu_{\text{ri}}} \left(\sum\limits_{i=N_{nt}+2}^{N_{nc}}{V_i}+\sum\limits_{i=N_{nc}+2}^{N_{n(t+h)}}{V_i}\right)-\frac{N_{n(t+h)}-N_{nt}-1}{nh/ \mu_{\text{ri}}}\E[V_i]\right)\frac{1}{s^{(n)}_t }.
\end{align*}
Thus, using the above arguments, we can conclude equation (\ref{eq:GammaToL}), and it only remains to be shown that the true means $\mu_1, \mu_2$ and the true scaling $s$ can be replaced by their estimators.  \\

 \noindent \underline{Step 2: replacement of parameters by their estimators}\\
First, we show equation (\ref{eq:hatGamma}). In order to obtain $\hat\Gamma^{(n)}$, we replace the true means by their estimators in the numerator of $\widetilde{\Gamma}^{(n)}$ and the true rate change point $c$ by $\hat c$.  
Our aim is to show
\begin{equation}
\label{eq:Step2}
\left(\frac{\sqrt{n}}{N_{n(t+h)}-N_{nt}} \left(\sum\limits_{i=N_{nt}+2}^{N_{n(t+h)}}{(\xi_i-\mu^{(i)})^2}-\sum\limits_{i=N_{nt}+2}^{N_{n(t+h)}}{(\xi_i-\hat\mu^{(i)})^2}\right)\right)_t  \xlongrightarrow[]{\mathbb{P}} (0)_t
\end{equation}
for the right window with analogous arguments for the left window. 

To simplify notation we now restrict to $(D(c-h,c],d_{||\cdot||})$ and show that (\ref{eq:Step2}) holds. The corresponding convergences in $(D[0,c-h],d_{||\cdot ||})$ and  $(D(c,T-h],d_{||\cdot ||})$ can be shown with similar arguments. 
For our notation we assume $(nt,nt+nh] \ni n\hat c$ where analogous arguments can be applied for the case $(nt,nt+nh] \niton  n\hat c$.
We first use the local estimators
\begin{align*}
\hat\mu_{1,\text{loc},t}:= (N_{nc}-N_{nt}-1)^{-1} \sum_{i=N_{nt}+2}^{N_{nc}}{\xi_i} \quad \text{and} \quad 
\hat\mu_{1,\text{loc},t}^{\hat c}:= (N_{n \hat c}-N_{nt}-1)^{-1} \sum_{i=N_{nt}+2}^{N_{n \hat c}}{\xi_i}
\end{align*} 
and analogously for $\hat\mu_{2,\text{loc}, t}, \hat\mu_{2,\text{loc}, t}^{\hat c}$. Applying the same arguments as for $R_t^{(n)}$ in the proof of Theorem \ref{main_theorem_a} we conclude that in $(D(c-h,c],d_{||\cdot||})$
\begin{equation}
\label{eq:knownbef}
\left(\frac{\sqrt{n}}{N_{n(t+h)}-N_{nt}} ((N_{nc}-N_{nt}) (\mu_1 - \hat \mu_{1,\text{loc}, t})^2 + (N_{n(t+h)}-N_{nc}) (\mu_2- \hat \mu_{2,\text{loc}, t})^2)\right)_t \xlongrightarrow[]{\mathbb{P}} (0)_t.
\end{equation}
Thus, we have to prove that the difference of (\ref{eq:knownbef}) and (\ref{eq:Step2}) vanishes, i.e.,
\begin{align}
\label{eq:tse}
\notag
&\left(\frac{\sqrt{n}}{N_{n(t+h)}-N_{nt}}\right)_t \times \\
\notag
&\left(((N_{n \hat c}-N_{nt}) {\left(\hat\mu_{1,\text{loc}, t}^{\hat c}\right)}^2 - (N_{nc}-N_{nt})  \hat \mu_{1,\text{loc}, t}^2 
 + (N_{n(t+h)}-N_{n \hat c}) {\left(\hat\mu_{2,\text{loc}, t}^{\hat c}\right)}^2 -  (N_{n(t+h)}-N_{nc}) \hat \mu_{2, \text{loc}, t}^2)\right)_t \\
& \quad\quad\quad \xlongrightarrow[]{\mathbb{P}} (0)_t.
\end{align}
We concentrate on the first two terms with the argumentation for the other terms being similar and note that the corresponding terms in the previous line are the same as
\begin{align}
\label{eq:same}
\frac{\sqrt{n}}{N_{n(t+h)}-N_{nt}} \left((N_{n \hat c}- N_{nc}){\left(\hat\mu_{1,\text{loc}, t}^{\hat c}\right)}^2 - (N_{nc}- N_{nt}) \left(\hat \mu_{1,\text{loc}, t}^2 - {\left(\hat\mu_{1,\text{loc}, t}^{\hat c}\right)}^2\right)\right).
\end{align}
Due to assumption (\ref{eq:assump}) we derive $|\sum_{i=N_{n \hat c}+1}^{N_{nc}}{\xi_i}| \leq |n(c - \hat c)| = o_{\mathbb{P}}(1)$. As the life times are positive, we consequently have for the number of summands
\begin{equation}
\label{eq:numbersummands}
N_{n \hat c} - N_{nc} \xlongrightarrow[]{\mathbb{P}} 0.
\end{equation}
This convergence does not depend on $t$ and consequently, it can be shown that the first summand in (\ref{eq:same}) vanishes. 
Defining $d_t^{(n)}:= \sqrt{n}(\hat \mu_{1,\text{loc}, t} - \hat \mu_{1,\text{loc}, t}^{\hat c})$ we now show that $(|\sqrt{n}(\hat \mu_{1,\text{loc}, t}^2 - {\left(\hat\mu_{1,\text{loc}, t}^{\hat c}\right)}^2)|)_t=(|2 \sqrt{n} \hat \mu_{1,\text{loc}, t}^{\hat c} d_t^{(n)} + {(d_t^{(n)})}^2|)_t$ converges in probability in $(D(c-h,c],d_{||\cdot||})$ to zero and thus (\ref{eq:tse}) holds. W.l.o.g. we assume $\hat{c} < c$ and observe
\begin{equation*}
\sqrt{n} d_t^{(n)}=n (N_{nc}-N_{nt})^{-1}\left(\sum\limits_{i=N_{nt}+2}^{N_{n \hat c}}{\xi_i}(N_{n \hat c}-N_{nt})^{-1} \left(N_{n \hat c}- N_{nc}\right) + \sum\limits_{i=N_{n \hat c}+1}^{N_{nc}}{\xi_i} \right).
\end{equation*}
Using Lemma \ref{lemma_nhprocess} and equation (\ref{eq:numbersummands}), we can prove that $(\sqrt{n} d_t^{(n)})_t$ (and also $({(d_t^{(n)})}^2)_t$)  vanishes in $(D(c-h,c],d_{||\cdot||})$.

Thus, applying Slutsky's theorem we have shown (\ref{eq:hatGamma}) for the locally estimated means $\hat \mu_{1,\text{loc}, t}, \hat \mu_{2,\text{loc}, t} $. The substitution of these locally estimated means by the global means $\hat \mu_1^{\hat c}:=(N_{n\hat c}-1)^{-1}\sum_{i=1}^{N_{n \hat c}}{\xi_i},\hat \mu_2^{\hat c}:=(N_{nT}-N_{n \hat c}-1)^{-1}\sum_{i=N_{n \hat c}+2}^{N_{nT}}{\xi_i}$ can be done with a decomposition argument similiar to the one in the proof of Theorem \ref{main_theorem_a}. Hence, an application of Slutsky's theorem allows us to finally show (\ref{eq:hatGamma}). 

In the last part of the proof we substitute the interpolated variance $(s_t^{(n)})^2$ by the estimated variance $\hat {s}^2_{nh,nt}$. With Lemma \ref{lemma_conv_s2} (section \ref{techn_lemmas}) and  weak convergence of $\hat {\Gamma} \xrightarrow[]{} \widetilde L$, the assertion follows.
$\Box$

\subsubsection{Technical Lemmas}		\label{techn_lemmas}
Using $\mu_{\text{le}}$ and $\mu_{\text{ri}}$ we obtain a convergence result for the scaled counting process $(N_t)_{t \geq 0}$.
\begin{lemma}
	\label{lemma_nhprocess}	
	Let $\Phi$ be a renewal process like in Theorem \ref{main_theorem_b} with mean functions $\mu_{\text{le},h,t}$, $\mu_{\text{ri},h,t}$ as in (\ref{def_muaverage2}). Let $T>0$, $h \in (0,T/2]$. Then we have in $(D[\tau_h],d_{||\cdot||})$ as $n \to \infty$ almost surely
	\begin{align*}
		\left(\frac{N_{n(t+h)}-N_{nt}}{nh / \mu_{\text{ri},h,t}}\right)_{t \in \tau_h}  \xlongrightarrow[]{}  (1)_{t \in \tau_h} \quad \text{and} \quad
		\left(\frac{N_{nt}-N_{n(t-h)}}{nh / \mu_{\text{le},h,t}}\right)_{t \in \tau_h}  \xlongrightarrow[]{}  (1)_{t \in \tau_h}.
	\end{align*}
\end{lemma}
\noindent
\textit{Proof}: This is Lemma A.1 in \citet{messer_shark}.

The next result shows the convergence of the denominator of $G$. 			
For a known rate the estimator $\hat \nu_{\text{le}}^2$ (\ref{def_nu_exact}) may be written as
\begin{equation}
\hat{\nu}^2_{\text{le}}:=\frac{1}{N_{nt}-N_{n(t-h)}-1}\left({\sum\limits_{i=N_{n(t-h)}+2}^{N_{nc}}{((\xi_i-\mu_1)^2-\hat{\sigma}_{\text{le}}^2)^2}+\sum\limits_{i=N_{nc}+2}^{N_{nt}}{((\xi_i-\mu_2)^2-\hat{\sigma}_{\text{le}}^2)^2}}\right)
\label{eq:defnu}
\end{equation}
and analogous for $\hat \nu_{\text{ri}}^2$ where $c$ denotes the rate change point. Note that the life time $\xi_{N_{nc}+1}$ is not considered in the terms above as its distribution is a mixture of two distributions.

\begin{lemma}
\label{lemma_conv_s2}
Let $\Phi_1(\mu_1,\sigma^2)$ and $\Phi_2(\mu_2,\sigma^2)$ be independent elements of $\mathscr{R}$  with $\mu_1 \neq \mu_2$.  Let $ c \in (0,T]$ be a rate change point, so that the sequence $\Phi^{(n)}$ results from $\Phi_1$ and $\Phi_2$ according to model (\ref{model_exp}). Let $\hat{s}_{nh,nt}$ and $s_t^{(n)}$be defined as in (\ref{def_s}) and (\ref{def_stheo}) and $\hat c$ be an estimator of $c$ fulfilling assumption (\ref{eq:assump}). Then it holds in $(D[\tau_h],d_{||\cdot||})$ for $n \to \infty$ in probability
\begin{equation*}
\left(\hat{s}_{nh,nt}\right)_{t \in \tau_h} \xlongrightarrow []{}  \left(s_t^{(1)}\right)_{t \in \tau_h} 
\end{equation*}
\end{lemma}

\noindent
\textit{Proof}: 
We show that the limit behaviour of $\hat{s}_{nh,nt}$ is given by
\begin{equation*}
\frac{{\nu}_{\text{ri}}^2}{h /\mu_{\text{ri}}}+\frac{{\nu}_{\text{le}}^2}{h /\mu_{\text{le}}} \quad \forall t \in \tau_h,
\end{equation*}
where $\mu_{\text{ri}}$ and $\mu_{\text{le}}$ are the window means defined in (\ref{def_muaverage2}). \\
 $\nu^2_{\text{ri}}:=\nu^2_{\text{ri}}(h,t)$ is given by ${\nu_1}^2$ for $t \leq c-h$, by  ${\nu_2}^2$ for $t > c$ and by
\begin{equation}
\label{def_nurl}
\nu^2_{\text{ri}}(h,t):=\frac{(c-t) /\mu_1 \cdot {\nu_1}^2 +(t+h-c) / \mu_2 \cdot {\nu_2}^2}{h /\mu_{\text{ri}}}
\end{equation}
otherwise. $\nu^2_{\text{le}}$ is defined analogously. \\
The uniform a.s. convergence of the estimators $\hat{\mu}_{\text{ri}}$ and $\hat{\mu}_{\text{le}}$ to $\mu_{\text{ri}}$ and $\mu_{\text{le}}$ is shown in Lemma A.2 in \citet{messer_shark}.  The uniform convergence in probability of the estimators $\hat{\nu}^2_{\text{ri}}$ and $\hat{\nu}^2_{\text{ri}}$ (as in (\ref{def_nu_exact})) to $\nu^2_{\text{ri}}$ and $\nu^2_{\text{le}}$ (defined in (\ref{def_nurl})) can be shown using the consistency result for $\nu^2$ (Lemma \ref{nu_kons}). We show this for $\hat{\nu}^2_{\text{le}}$ with the argumentation for $\hat{\nu}^2_{\text{ri}}$ being similar and assume first a known mean profile. By Lemma \ref{nu_kons} it holds for $n \to \infty$ that
\begin{align*}
\left(\frac{\sum\limits_{i=N_{n(t-h)}+2}^{N_{nc}}{((\xi_i-\mu_1)^2-\hat{\sigma}_{\text{le}}^2)^2}}{N_{nc}-N_{n(t-h)}-1}\right)_{t \in \tau_h} \xlongrightarrow[]{a.s.} (\nu^2_1)_{t \in \tau_h}.
\end{align*}
Lemma \ref{lemma_nhprocess} and Slutsky's theorem imply for the first summand of (\ref{eq:defnu})
\begin{align*}
\left(\frac{\sum\limits_{i=N_{nt-nh}+2}^{N_{nc}}{((\xi_i-\mu_1)^2-\hat{\sigma}_{\text{le}}^2)^2}}{N_{nt}-N_{n(t-h)}-1}\right)_{t \in \tau_h} \xlongrightarrow[]{a.s.}\left(\frac{(c-(t-h))/ \mu_1}{h / \mu_{\text{ri}}}\nu^2_1\right)_{t \in \tau_h}.
\end{align*}
Similar calculations for the second summand yield 
\begin{align*}
\left(\frac{\sum\limits_{i=N_{nc}+2}^{N_{nt}}{((\xi_i-\mu_2)^2-\hat{\sigma}_{\text{le}}^2)^2}}{N_{nt}-N_{n(t-h)}-1}\right)_{t \in \tau_h} \xlongrightarrow[]{a.s.} \left(\frac{(t-c)/ \mu_2}{h / \mu_{\text{ri}}}\nu^2_2\right)_{t \in \tau_h}.
\end{align*}
The  exchange of the true means by their estimators results from Slutsky's theorem using  assumption (\ref{eq:assump}) of consistency (in probability) of the change point estimator, which yields $(\hat{\nu}^2_{\text{le}})_{t \in \tau_h} \xlongrightarrow []{} (\nu^2_{\text{le}})_{t \in \tau_h}$ in probability. As all four functions $(\mu_{\text{ri}})_{t \in \tau_h}, (\mu_{\text{le}})_{t \in \tau_h}, (\nu^2_{\text{ri}})_{t \in \tau_h}$ and $(\nu^2_{\text{le}})_{t \in \tau_h}$ are continuous $\hat{s} \xrightarrow []{} s^{(1)}$ holds in probability for $n \to \infty$ by the form of the estimator $\hat{s}$ in (\ref{def_s}).
$\Box$ \\
Note that Theorem \ref{main_theorem_b} holds not only for renewal processes but for all point processes for which (\ref{eq:donsker}) holds and $s_t$ is consistently estimated, for example also for a subclass of renewal processes with varying variance (RPVVs, see~\citet{messer_paper}).

\end{appendix}

 \bibliography{literatur_paper}

\end{document}